\RequirePackage{fix-cm}
\documentclass[letterpaper,journal]{IEEEtran}
\usepackage{amsmath,amsfonts}
\usepackage{mathtools,nccmath}
\usepackage{algorithmic}
\usepackage{algorithm}
\usepackage{array}
\usepackage{tikz}
\usetikzlibrary{arrows.meta}
\usepackage{ctable}
\usepackage[caption=false,font=normalsize,labelfont=sf,textfont=sf]{subfig}
\usepackage{textcomp}
\usepackage{stfloats}
\usepackage{url}
\usepackage{verbatim}
\usepackage{graphicx}
\graphicspath{{figures/}}
\usepackage{cite}
\usepackage{multirow}
\usepackage{booktabs}
\makeatletter
\newcommand{\insettablerules}{%
  \def\@BTnormal{%
    {\CT@arc@\nointerlineskip
      \hbox to\linewidth{\kern\cmidrulekern
        \leaders\hrule height\@thisrulewidth\hfill
        \kern\cmidrulekern}%
      \nointerlineskip}%
    \futurenonspacelet\@tempa\@BTendrule}}
\makeatother
\usepackage{makecell}
\DeclareMathOperator*{\argmax}{argmax}
\def\BibTeX{{\rm B\kern-.05em{\sc i\kern-.025em b}\kern-.08em
    T\kern-.1667em\lower.7ex\hbox{E}\kern-.125emX}}
\usepackage{balance}
\usepackage[protrusion=false,expansion=true]{microtype}
\usepackage[dont-mess-around]{fnpct}
\usepackage{bigfoot}
\def\IEEEbibitemsep{-0.35pt plus 0.22pt minus 0.15pt}
\makeatletter
\def\section{\@startsection{section}{1}{\z@}{1.60ex plus 0.42ex minus 0.32ex}%
  {0.40ex plus 0.18ex minus 0ex}{\normalfont\normalsize\centering\scshape}}
\def\subsection{\@startsection{subsection}{2}{\z@}{1.08ex plus 0.32ex minus 0.22ex}%
  {0.26ex plus 0.15ex minus 0ex}{\normalfont\normalsize\itshape}}
\makeatother
\AtBeginDocument{%
  \setlength{\abovedisplayskip}{4.9pt plus 1.1pt minus 1.0pt}%
  \setlength{\belowdisplayskip}{4.9pt plus 1.1pt minus 1.0pt}%
  \setlength{\abovedisplayshortskip}{2.8pt plus 0.8pt minus 0.8pt}%
  \setlength{\belowdisplayshortskip}{3.8pt plus 0.8pt minus 0.8pt}%
  \setlength{\parskip}{0pt}%
}
\begin{document}
\bstctlcite{CAST:BSTcontrol}

\title{BiSAT: Bidirectionally Aligned Semantic-Acoustic\\ Tokenization for Speech Language Models}

\author{Yixian~Lu\textsuperscript{*}, Lijun~Wang\textsuperscript{*},~\IEEEmembership{Graduate~Student~Member,~IEEE,} and~Shogo~Okada\textsuperscript{$\dagger$},~\IEEEmembership{Member,~IEEE}%
\thanks{The authors are with the Japan Advanced Institute of Science and Technology, Nomi, Ishikawa 923-1292, Japan (e-mail: \mbox{yixian.lu@jaist.ac.jp}; \mbox{lijun.wang@jaist.ac.jp}; \mbox{okada-s@jaist.ac.jp}). This work was partially supported by JSPS KAKENHI (No.\ 26K03016), JST CREST (No.\ JPMJCR2563), JST CRONOS (No.\ JPMJCS24K7), and AMED under Grant Number JP256f0137001.}%
\thanks{* Equal contribution; author order was randomly determined by tossing a coin. $\dagger$ Corresponding author.}}

\markboth{B\MakeLowercase{i}SAT: Bidirectionally Aligned Semantic-Acoustic Tokenization}%
{Lu \MakeLowercase{\textit{et al.}}: B\MakeLowercase{i}SAT for Speech Language Models}


\IEEEaftertitletext{\vspace{-1.95\baselineskip}}
\maketitle

\begin{abstract}
Speech language models (LMs) generate speech as sequences of discrete tokens, requiring tokenizers that preserve both acoustic detail and linguistic content at a low bitrate. Semantic-enhanced tokenizers may discard acoustic detail, whereas acoustic-oriented tokenizers retain acoustic variation that hinders language modeling. Unified tokenizers combine both, but encoding them within a single feature space can cause interference, and existing methods typically align acoustic features to fixed targets from a frozen semantic teacher. We propose Bidirectionally Aligned Semantic-Acoustic Tokenization (BiSAT), a unified framework with separate acoustic and semantic branches coupled by a shared-index mechanism. During training, each branch has its own encoder, codebook, and decoder, and at every frame a shared index minimizes the combined quantization error across both branches. BiSAT thus learns acoustic and semantic representations in separate continuous spaces while producing a single discrete code stream, jointly optimizing both encoders without fixed semantic teachers. Built on BiSAT, BiSAT-TTS pairs an autoregressive text-speech LM with a conditional flow-matching acoustic decoder for zero-shot text-to-speech. Reconstruction, frozen-code understanding, and diagnostic analyses show that BiSAT preserves accessible linguistic and acoustic structure in a low-bitrate single stream, generalizes strongly to in-the-wild and out-of-domain speech, and learns complementary paired codebooks. BiSAT-TTS matches or surpasses state-of-the-art speech LM-based TTS baselines with less training data, demonstrating that accessible code structures and patterns, not only reconstruction quality alone, benefits autoregressive speech modeling. The audio demos are available at \textcolor{blue}{\url{https://seekmind-lab.github.io/bisat-demo}}.
\end{abstract}


\section{Introduction}
\label{sec:introduction}
\IEEEPARstart{S}{peech} language models (LMs) generate speech as sequences of discrete tokens, enabling text-to-speech (TTS) systems to adopt large language model (LLM) architectures and training recipes for scalable sequence modeling \cite{wang2023neural,du2024cosyvoice2,ye2025llasa}. In particular, autoregressive (AR) speech LMs determine output length flexibly without explicit duration modeling, in contrast to many non-autoregressive (NAR) methods \cite{wang2023neural}. Before a token-based LM can model speech, the continuous speech waveform must be discretized. This can be performed by quantization, such as vector quantization (VQ) \cite{vandenoord2017vqvae}, residual vector quantization (RVQ) \cite{zeghidour2022soundstream}, finite scalar quantization (FSQ) \cite{mentzer2024fsq}, and binary spherical quantization (BSQ) \cite{zhao2024bsq}. VQ, FSQ, and BSQ give one code per frame, whereas RVQ quantizes the residual of each stage with a further codebook and produces several code streams.
Following common usage in speech tokenization \cite{borsos2023audiolm}, we use ``\emph{semantic}'' to refer to the linguistic content of speech, including phonetic and lexical information. The training objective influences what information the codes retain \cite{vandenoord2017vqvae}. Speech reconstruction encourages the codes to retain acoustic information such as prosody, pitch, timbre, and speaker identity \cite{zeghidour2022soundstream,defossez2023encodec,ju2024naturalspeech3,wu2024codecsuperb}, whereas speech LMs also require codes that preserve semantic content. Retaining both types of information within a compact single-stream bottleneck is a central trade-off in designing a tokenizer for a speech LM. The information retained determines what the LM must model and what the acoustic decoder can recover to resynthesize the waveform.\looseness=-1

In general, LM-based TTS uses three types of speech tokenizers that differ in how they make this trade-off. \emph{Semantic-enhanced} tokenizers (SETs) insert a quantizer into a pretrained speech understanding model \cite{du2024cosyvoice, du2024cosyvoice2, du2025cosyvoice3, wang2025sparktts, hu2026qwen3tts}. \emph{Acoustic-oriented} tokenizers (AOTs) learn codes primarily through waveform or mel-spectrogram reconstruction \cite{zeghidour2022soundstream,defossez2023encodec,dellalibera2025focalcodec,ji2025wavtokenizer}. \emph{Unified} tokenizers (UTs) combine acoustic reconstruction with semantic supervision \cite{zhang2024speechtokenizer,defossez2024moshi,gong2025xytokenizer,ye2025xcodec,jo2026lmspt,ye2025llasa}. Semantic supervision captures linguistic content but may discard fine acoustic details needed for reconstruction. Conversely, codes learned through reconstruction preserve acoustic detail well, yet they also retain variation in timbre, prosody, and recording conditions that can weaken alignment with text and cause AR LMs to repeat or skip words \cite{zhang2024speechtokenizer,chen2024valle2,song2024ellav}. Recent UTs combine acoustic and semantic supervision in one tokenizer, either by guiding the first RVQ layer with semantic features \cite{zhang2024speechtokenizer,defossez2024moshi,jo2026lmspt} or by fusing the two representations before quantization \cite{gong2025xytokenizer,ye2025xcodec,ye2025llasa}.
However, encoding acoustic detail and linguistic content within a single shared or fused representation space can cause interference between the two objectives. Furthermore, these methods derive semantic supervision from pretrained or frozen encoders, so the semantic targets remain fixed during tokenizer training.
Finally, most AOTs and UTs reconstruct waveforms with adversarial training (AT), since regression losses alone yield over-smoothed speech at low bitrates \cite{zeghidour2022soundstream,defossez2023encodec}. Because an adversarial discriminator evaluates naturalness rather than code faithfulness, a co-trained waveform decoder can synthesize details absent from the codes, decoupling reconstruction quality from structured semantic patterns \cite{defossez2023encodec,ji2025wavtokenizer,zhang2024speechtokenizer,ye2025llasa}, as discussed in Section\,\ref{ssec:work_tokenization}.\looseness=-1

These limitations motivate us to design a low-bitrate single-stream tokenizer at a low frame rate that learns acoustic and semantic representations in separate continuous spaces while jointly optimizing both branches without fixed semantic teachers. TokenFlow \cite{qu2024tokenflow} showed in image tokenization that separate semantic and pixel-level codebooks linked by shared indices can support both understanding and generation. Motivated by this shared-index principle, we propose Bidirectionally Aligned Semantic-Acoustic Tokenization (BiSAT). During training, BiSAT uses separate acoustic and semantic branches, each with its own encoder, codebook, and decoder. The acoustic branch reconstructs mel spectrograms without adversarial training, and the semantic branch is trained with connectionist temporal classification (CTC). Waveform synthesis is deferred to a separate acoustic decoder and vocoder during TTS, so the tokenizer codes are shaped solely by mel reconstruction and CTC supervision. At each frame, a shared-index vector quantizer selects a single index by minimizing the weighted sum of squared Euclidean distances across both branches. This design offers two key advantages. First, under a finite discrete bottleneck \cite{tishby1999information, alemi2017deep}, keeping the encoders, codebooks, and training decoders in separate continuous spaces is intended to reduce potential interference between acoustic reconstruction and semantic recognition, while allowing each branch to learn task-specific representations. Second, unlike approaches that align acoustic features to fixed semantic teacher targets, BiSAT jointly optimizes both encoders. Since the shared index depends on both branches, updates to either encoder influence the quantization assignments of both branches. We refer to this coupled optimization mechanism as bidirectional alignment. We evaluate whether the resulting single code stream retains information useful for both linguistic and acoustic downstream tasks.\looseness=-1

We evaluate BiSAT on speech reconstruction from codes and on SUPERB and ML-SUPERB understanding probes \cite{yang2021superb,shi2023mlsuperb} across linguistic, speaker, and paralinguistic tasks to verify its effectiveness in preserving accessible semantic content and acoustic detail for speech LMs. We also analyze the paired codebooks and evaluate zero-shot TTS with BiSAT-TTS, combining an AR text-speech LM, a reference-conditioned flow-matching acoustic decoder, and a pretrained vocoder.\looseness=-1

Our main contributions are summarized as follows:
{\setlength{\IEEEiedtopsep}{0pt plus 0.5pt}%
\begin{enumerate}
  \item We propose BiSAT, a unified speech tokenization framework that learns separate continuous acoustic and semantic representations while coupling index assignment across paired codebooks to produce a single low-bitrate code stream.\looseness=-1
  \item We develop BiSAT-TTS, a zero-shot TTS system that combines an AR text-speech LM with a conditional flow-matching acoustic decoder using BiSAT tokens.
  \item We provide a comprehensive evaluation of BiSAT across speech reconstruction, downstream understanding, diagnostic analysis, and zero-shot TTS, showing competitive or superior performance across multiple benchmarks and validating the effectiveness of the overall framework.
\end{enumerate}}

The rest of the manuscript is organized as follows. Section\,\ref{sec:related} reviews related work. Section\,\ref{sec:method} describes BiSAT and BiSAT-TTS. Section\,\ref{sec:exp} details experiments. Section\,\ref{sec:conclusion} concludes this work and Section\,\ref{sec:limitation} discusses limitations and future work.

\section{Related Work}
\label{sec:related}

\subsection{Speech Tokenization for Speech Language Modeling}
\label{ssec:work_tokenization}
\subsubsection{Semantic-Enhanced Tokenizers}
\label{sssec:work_semantic_tok}
SETs insert quantization modules into pretrained speech understanding models. The CosyVoice series \cite{du2024cosyvoice,du2024cosyvoice2,du2025cosyvoice3} quantizes intermediate representations of an understanding model, progressing from VQ with automatic speech recognition (ASR) supervision to FSQ and, in CosyVoice 3, multi-task supervision with ASR, language identification, emotion recognition, audio event detection, and speaker analysis. Spark-TTS \cite{wang2025sparktts} combines frame-level semantic tokens with fixed-length global tokens for utterance-level properties, and Qwen3-TTS \cite{hu2026qwen3tts} uses a 25\,Hz single-codebook tokenizer that emphasizes semantic content. The CosyVoice tokenizers are trained only with understanding tasks, so their tokens carry little of the acoustic detail needed for reconstruction. Flow-based \cite{du2024cosyvoice2,du2025cosyvoice3} or generative adversarial network (GAN) decoders \cite{wang2025sparktts} are thus employed to recover this detail from additional conditioning, which may demand more data and larger decoder capacity.

\subsubsection{Acoustic-Oriented Tokenizers}
\label{sssec:work_acoustic_tok}
AOTs learn codes through reconstruction. RVQ codecs such as EnCodec \cite{defossez2023encodec} add residual quantization layers to recover finer acoustic detail, but serializing multiple code streams into one sequence significantly increases the sequence length for AR modeling. The VALL-E series \cite{wang2023neural,chen2024valle2} therefore predicts the first stream with an AR LM and the residual streams with a parallel NAR model. VoiceCraft and VoiceStar \cite{peng2024voicecraft,peng2025voicestar} predict all streams in one AR LM through a delayed codebook pattern, and CLaM-TTS \cite{kim2024clamtts} models probabilistic RVQ latents. WavTokenizer \cite{ji2025wavtokenizer} and FocalCodec \cite{dellalibera2025focalcodec} reduce the sequence burden for language modeling by producing a single code stream at a low bitrate. We also adopt a single low-bitrate code stream at a low frame rate for efficient AR language modeling without predicting additional residual streams.

These AOTs receive no explicit semantic supervision, so the LM must infer linguistic structure from acoustically grounded tokens, which can cause repetitions, skipped words, and non-terminating sequences \cite{chen2024valle2,song2024ellav,song2024tacolm}. Another problem comes from coupling tokenization with waveform reconstruction. Most of these codecs reconstruct waveforms with AT, since regression losses alone produce over-smoothed speech at low bitrates \cite{zeghidour2022soundstream,defossez2023encodec}. The discriminator judges whether the output sounds natural, not whether it matches the input. The decoder can thus synthesize details that the codes do not contain, and the adversarial objective puts no direct pressure on the codes to form structured semantic patterns. This makes language modeling harder and demands more data, model capacity, and computation. These observations motivate us to decouple tokenization from waveform reconstruction and to supervise the tokenizer with semantic content directly.

\subsubsection{Unified Tokenizers}
\label{sssec:work_unified_tok}
UTs combine semantic supervision with acoustic reconstruction. SpeechTokenizer \cite{zhang2024speechtokenizer} and Mimi \cite{defossez2024moshi} guide the first RVQ layer with self-supervised learning (SSL) features from HuBERT or WavLM, while later layers encode the residual acoustic information. LM-SPT \cite{jo2026lmspt} reconstructs speech from its semantic tokens and matches features of the original and reconstructed speech extracted by a frozen speech encoder. In these RVQ designs, the first layer must represent linguistic content while providing the acoustic approximation refined by later layers, which makes the two objectives hard to balance \cite{zhang2024speechtokenizer,defossez2024moshi,jo2026lmspt,hussein25_interspeech}. XY-Tokenizer \cite{gong2025xytokenizer} and the X-Codec series \cite{ye2025xcodec,ye2025llasa} instead fuse the outputs of acoustic and semantic encoders before quantization. X-Codec2 quantizes the fused features with a single FSQ codebook, which allows LLaSA \cite{ye2025llasa} to use one AR LM. In fusion designs, encoding acoustic detail and linguistic content within a single fused representation can cause interference between the two objectives during training \cite{gong2025xytokenizer}. These methods also derive semantic supervision from pretrained or frozen speech encoders, namely teacher features \cite{zhang2024speechtokenizer,defossez2024moshi,ye2025xcodec,ye2025llasa} or a frozen encoder \cite{gong2025xytokenizer,jo2026lmspt}, so the semantic targets remain fixed during tokenizer training. Finally, UTs couple tokenization with adversarial waveform reconstruction as in AOTs, allowing reconstruction to rely on co-trained decoders without ensuring generalizable semantic code patterns.

In image tokenization, TokenFlow \cite{qu2024tokenflow} uses separate semantic and pixel-level codebooks linked by shared indices for understanding and generation. Inspired by this design, BiSAT jointly updates acoustic and semantic encoders with separate codebooks and determines their shared discrete indices from both representations.\looseness=-1

\subsection{Zero-shot TTS Systems}
\label{ssec:work_zs_tts}
Zero-shot TTS systems include NAR, AR, continuous AR, and discrete diffusion models. NAR systems generate all frames in parallel, including flow- and VAE-based models with monotonic alignment search or explicit duration modeling \cite{mehta2024matcha,lee2025hierspeechpp,yang2025simplespeech2}, latent diffusion models \cite{shen2024naturalspeech2,ju2024naturalspeech3}, and flow-matching models that learn alignment without supervision \cite{eskimez2024e2tts,chen2025f5tts}. Continuous AR models predict continuous acoustic features directly, which avoids quantization error but can suffer from oversmoothing and unstable sequential generation \cite{meng2025autoregressive,zhou2025voxcpm}. Discrete diffusion models instead generate token sequences through iterative denoising \cite{moon2026deltatts}. Discrete AR text-speech LMs generate speech tokens sequentially without explicit duration control. While RVQ-based systems require a multi-stage or parallel prediction scheme \cite{wang2023neural,chen2024valle2,kim2024clamtts}, single-stream VQ or FSQ tokenizers allow a LM \cite{du2024cosyvoice,du2024cosyvoice2,wang2025sparktts,ye2025llasa}. BiSAT-TTS follows this single-stream AR approach.

\subsection{Acoustic-Semantic Representation Alignment}
\label{ssec:work_prh}
In speech, representation-level alignment between acoustic signals and linguistic content has been studied through self-supervised pretraining \cite{baevski2020wav2vec2,hsu2021hubert,chen2022wavlm}, joint speech-text pretraining \cite{ao2022speecht5,tang2022unified}, and semantic regularization of generative latent spaces \cite{niu2026semanticvae}. These methods align acoustic and linguistic information within a single continuous space. BiSAT instead initializes dual encoders from a shared self-supervised speech backbone \cite{kuchaiev2019nemo}, specializes them under separate acoustic and semantic supervision, and couples the two continuous spaces through shared-index quantization into a single code stream.
\begin{figure*}[t]
    \centering
    \includegraphics[width=0.99\linewidth]{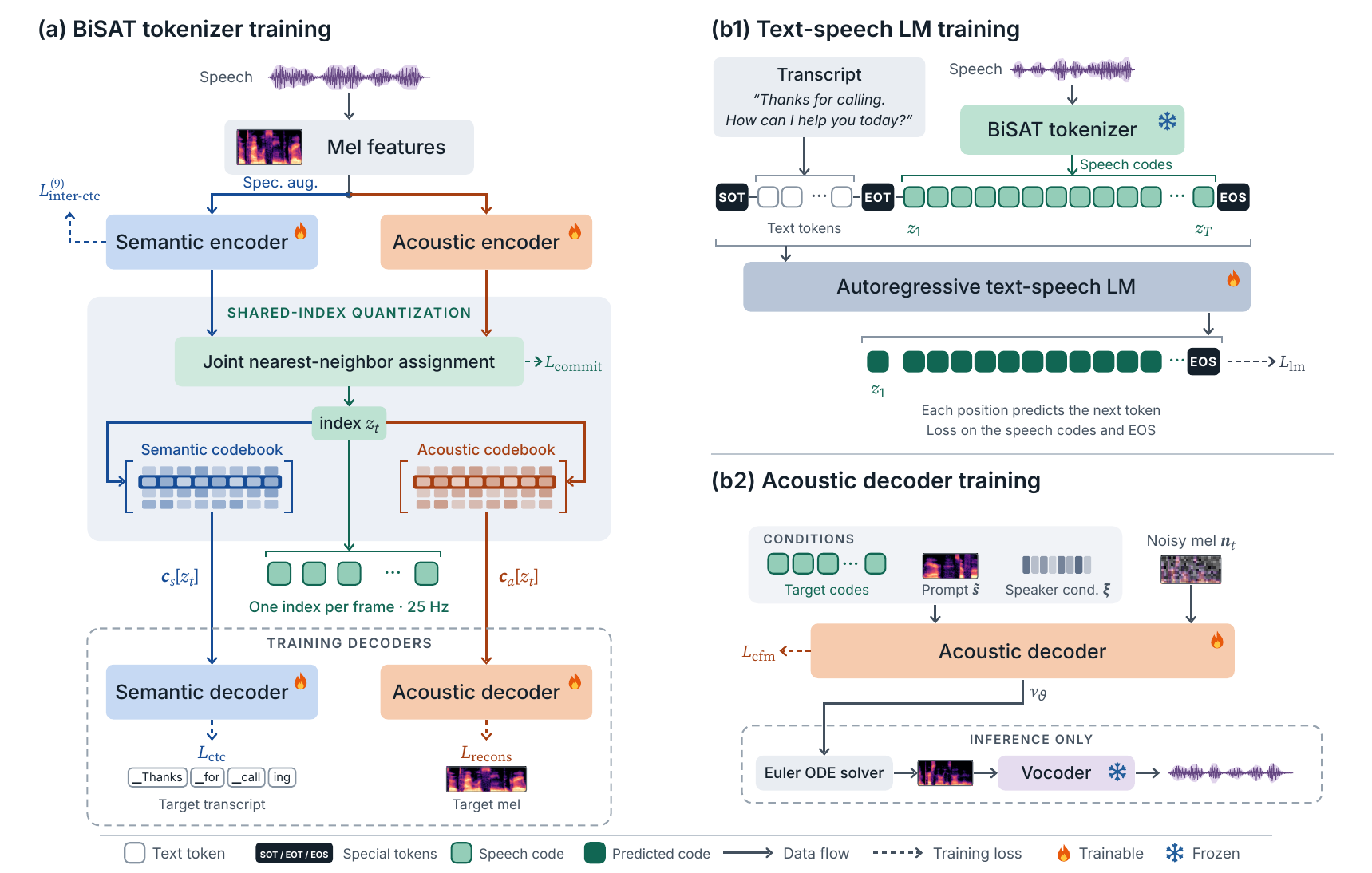}
    \vspace{-1.15em}
    \caption{Training of the BiSAT tokenizer and BiSAT-TTS. (a) The acoustic and semantic branches are jointly trained with mel-spectrogram reconstruction and CTC, respectively. A shared-index vector quantizer selects one entry per frame from each codebook to create a single code stream. A commitment loss is used to keep both encoder outputs close to these entries. (b1) The text-speech LM predicts each speech code and the final EOS from the text tokens and the preceding codes. (b2) The acoustic decoder predicts the flow-matching velocity from the noisy interpolant, conditioned on the codes, a prompt mel spectrogram, and a speaker condition. At inference, an Euler ODE solver generates the mel spectrogram, and the frozen vocoder converts it into a waveform.}
    \label{fig:overview}
\end{figure*}

\section{Methodology}
\label{sec:method}
\textbf{Overview.} Figure\,\ref{fig:overview} shows the BiSAT tokenizer and the BiSAT-TTS pipeline. The tokenizer (see Section\,\ref{sssec:cast_tokenizer}) discretizes a speech waveform into a single code sequence by aligning acoustic and semantic representations bidirectionally through shared-index quantization. For TTS, the text-speech LM (see Section\,\ref{sssec:cast_neural_codec_lm}) generates codes conditioned on text and reference speech. The flow-matching acoustic decoder (see Section\,\ref{sssec:cast_mel_decoder}) converts these codes to a mel spectrogram. A pretrained vocoder then synthesizes the waveform.

\subsection{Problem Formulation}
\label{ssec:prob_form}
We generate speech through an autoregressive text-speech LM \cite{wang2023neural,du2024cosyvoice2}. Let $(x,\boldsymbol{y})$ denote a speech-text pair, where $x$ is the waveform and $\boldsymbol{y}:=(y_1,\dots,y_U)$ is its tokenized transcript. CTC supervision and text-speech LM training use separate text tokenizers, with $\boldsymbol{y}$ and $U$ denoting the token sequence and its length in each setting. Let $\boldsymbol{m}\in\mathbb{R}^{F\times T_{\text{tok}}}$ denote the $F$-bin log-mel spectrogram encoded and reconstructed by the tokenizer, and let $\boldsymbol{s}\in\mathbb{R}^{F\times T_{\text{mel}}}$ denote the mel spectrogram generated by the TTS acoustic decoder. The two spectrograms are computed at different sampling and frame rates, as detailed in Section\,\ref{ssec:acoustic_features}. The BiSAT tokenizer maps $x$ to a single code sequence $\boldsymbol{z}_{1:T}=(z_1,\dots,z_T)$, where each code $z_t\in\{1,\dots,K\}$ corresponds to a shared index that selects one entry from each of two paired codebooks $\boldsymbol{c}_a,\boldsymbol{c}_s\in\mathbb{R}^{K\times D_{\text{codebook}}}$. Throughout the manuscript, subscripts $a$ and $s$ denote the acoustic and semantic branches, and $[\,\cdot\,;\,\cdot\,]^{(\cdot)}$ denotes concatenation along the indicated axis.

\subsubsection{BiSAT Tokenization}
\label{sssec:cast_tok_prob_form}
The tokenizer in Fig.\,\ref{fig:overview} learns acoustic and semantic representations aligned by shared discrete codes. Two encoders $\mathcal{E}_a$ and $\mathcal{E}_s$, parameterized by $\psi_a$ and $\psi_s$, encode $\boldsymbol{m}$ to time-aligned acoustic and semantic latents $\boldsymbol{e}_{a;1:T}$ and $\boldsymbol{e}_{s;1:T}$. Both encoders use the same feature dimension $D_a=D_s=D_{\text{codebook}}$. A shared-index quantizer converts these representations to paired quantized sequences $\hat{\boldsymbol{z}}_{a;1:T}$ and $\hat{\boldsymbol{z}}_{s;1:T}$. The mel reconstruction decoder $\mathcal{D}_a$, parameterized by $\gamma_a$, reconstructs the mel spectrogram $\hat{\boldsymbol{m}}\in\mathbb{R}^{F\times T_{\text{tok}}}$ from $\hat{\boldsymbol{z}}_{a;1:T}$. The semantic decoder $\mathcal{D}_s$, parameterized by $\gamma_s$, predicts semantic labels from $\hat{\boldsymbol{z}}_{s;1:T}$ and is trained with CTC supervision \cite{graves2006ctc}.\looseness=-1

The mel spectrogram $\boldsymbol{m}$ and text sequence $\boldsymbol{y}$ provide acoustic and semantic supervision for the shared codes $\boldsymbol{z}_{1:T}$. Following the VQ-VAE formulation \cite{vandenoord2017vqvae}, we define the audio-conditioned approximate posterior $q_\psi(\boldsymbol{z}_{1:T}\mid \boldsymbol{m})$ with $\psi=\{\psi_a,\psi_s\}$ as a deterministic nearest-neighbor assignment. Under a uniform prior over the $K$ codebook indices at each frame, the Kullback-Leibler (KL) term in the evidence lower bound (ELBO) on $\log p(\boldsymbol{m},\boldsymbol{y})$ equals the constant $T\log K$ and does not depend on the model parameters $\Theta=\{\psi_a,\psi_s,\gamma_a,\gamma_s,\boldsymbol{c}_a,\boldsymbol{c}_s\}$. Assuming that $\boldsymbol{m}$ and $\boldsymbol{y}$ are conditionally independent given $\boldsymbol{z}_{1:T}$, maximizing the ELBO reduces to:\looseness=-1
\begin{equation}
\label{eq:elbo_vae}
\begin{aligned}
  \argmax_{\Theta}\mathrm{ELBO}_{\text{tok}}
  &= \argmax_{\Theta}\mathbb{E}_{q_\psi(\boldsymbol{z}_{1:T}\mid\boldsymbol{m})}\!\big[\log p_{\gamma_a}(\boldsymbol{m}\mid\boldsymbol{z}_{1:T})\\
  &\qquad + \log p_{\gamma_s}(\boldsymbol{y}\mid\boldsymbol{z}_{1:T})\big].
\end{aligned}
\end{equation}
The former term can be simplified to an $\ell_1$ mel-spectrogram reconstruction loss for acoustic supervision, under a fixed-scale factorized Laplace likelihood. For the latter, CTC is employed to sum over frame-level label paths $\pi$ that collapse to $\boldsymbol{y}$ for semantic supervision. Specifically, the reconstruction and CTC losses $\mathcal{L}_{\text{recons}}$ and $\mathcal{L}_{\text{ctc}}$ are defined as:
\begin{equation}
\label{eq:rec_term}
\begin{aligned}
  & \mathcal{L}_{\text{recons}} = \frac{1}{T_{\text{tok}} F}\,\lVert \boldsymbol{m} - \hat{\boldsymbol{m}} \rVert_1, \\
  & \mathcal{L}_{\text{ctc}} = -\frac{1}{U}\,\log \sum_{\pi \in \mathcal{B}^{-1}(\boldsymbol{y})} \prod_{t=1}^{T} p_{\gamma_s}(\pi_t \mid \hat{\boldsymbol{z}}_{s;1:T}),
\end{aligned}
\end{equation}
where $\mathcal{B}^{-1}(\boldsymbol{y})$ is the set of all possible paths that collapse to $\boldsymbol{y}$, and $p_{\gamma_s}(\pi_t\mid\hat{\boldsymbol{z}}_{s;1:T})$ is the predicted probability of label $\pi_t$ at frame $t$. Since CTC marginalizes over monotonic alignments between code frames and text, no forced alignment is required.

\textbf{Intermediate CTC Supervision.} Following OWSM-CTC \cite{peng2024owsmctc}, we apply an auxiliary CTC head to the output $\boldsymbol{h}_s^{(9)}$ of the ninth semantic encoder layer. The loss is formulated as:
\begin{equation}
\label{eq:interctc_loss}
  \mathcal{L}_{\text{inter-ctc}}
  = -\frac{1}{U}\log \sum_{\pi \in \mathcal{B}^{-1}(\boldsymbol{y})} \prod_{t=1}^{T} p_{\gamma_{\text{int}}}
    \big(\pi_t \mid \boldsymbol{h}_s^{(9)}\big),
\end{equation}
where $p_{\gamma_{\text{int}}}$ is the label distribution of the auxiliary CTC head parameterized by $\gamma_{\text{int}}$.

\textbf{Shared-index Quantization.} Both encoder outputs are $\ell_2$-normalized frame by frame, i.e., $\bar{\boldsymbol{e}}_{b,t}=\boldsymbol{e}_{b,t}/\max(\lVert\boldsymbol{e}_{b,t}\rVert_2,\delta)$ for $b\in\{a,s\}$, with a small constant $\delta>0$ for numerical stability. The codebook entries $\{\boldsymbol{c}_{k,a},\boldsymbol{c}_{k,s}\}_{k=1}^K$ are also unit-normalized. The entries $\boldsymbol{c}_{k,a}$ and $\boldsymbol{c}_{k,s}$ form a pair at each index $k$. For each frame, the quantizer assigns a single index by minimizing the joint distance:
\begin{equation}
\label{eq:shared_map}
  z_t = \underset{k}{\arg\min} \left[ \lVert \bar{\boldsymbol{e}}_{a,t} - \boldsymbol{c}_{k,a} \rVert_2^2 + \omega \lVert \bar{\boldsymbol{e}}_{s,t} - \boldsymbol{c}_{k,s} \rVert_2^2 \right].
\end{equation}
Both branches retrieve the corresponding row as $\hat{\boldsymbol{z}}_{a,t} = \boldsymbol{c}_{z_t,a}$ and $\hat{\boldsymbol{z}}_{s,t} = \boldsymbol{c}_{z_t,s}$. For unit-normalized codebook entries, Eq.\,(\ref{eq:shared_map}) is equivalent to maximizing $\langle \bar{\boldsymbol{e}}_{a,t}, \boldsymbol{c}_{k,a} \rangle + \omega \langle \bar{\boldsymbol{e}}_{s,t}, \boldsymbol{c}_{k,s} \rangle$, so setting $\omega = 1$ weights the two branches equally. Each frame is represented by one index into $K$ code pairs. For example, a 25\,Hz code stream with $K=32768$ has a nominal bitrate of $25\log_2 32768=375$\,bps, regardless of $D_{\text{codebook}}$.

We use the rotation trick \cite{fifty2025rotation} to backpropagate the decoder gradients through quantization. For branch $b$ and frame $t$, let $\mathbf{R}_{b,t}$ be the rotation matrix that rotates the encoder output $\bar{\boldsymbol{e}}_{b,t}$ toward the selected codebook entry. The decoder input is parameterized as:
\begin{equation}
\label{eq:rotation_trick}
  \boldsymbol{q}_{b,t}
  = \mathbf{sg}\!\left(
      \frac{\lVert\hat{\boldsymbol{z}}_{b,t}\rVert_2}
           {\lVert\bar{\boldsymbol{e}}_{b,t}\rVert_2}
      \mathbf{R}_{b,t}\right)\bar{\boldsymbol{e}}_{b,t},
\end{equation}
where $\mathbf{sg}(\cdot)$ denotes the stop-gradient operator, so that $\boldsymbol{q}_{b,t}=\hat{\boldsymbol{z}}_{b,t}$ in the forward pass. Because both the encoder outputs and codebook entries are unit-normalized, the norm ratio equals one and the decoder gradient is multiplied by $\mathbf{R}_{b,t}^{\top}$ during backpropagation to the encoder.

We keep encoder outputs close to their assigned codebook entries by employing the commitment loss:
\begin{equation}
\label{eq:vq_term}
  \mathcal{L}_{\text{commit}} = \frac{1}{T}\!\sum_{t=1}^{T}\!\big[\lVert \bar{\boldsymbol{e}}_{a,t} - \mathbf{sg}(\hat{\boldsymbol{z}}_{a,t}) \rVert_2^2 + \lVert \bar{\boldsymbol{e}}_{s,t} - \mathbf{sg}(\hat{\boldsymbol{z}}_{s,t}) \rVert_2^2\big].
\end{equation}
The codebooks are updated using exponential moving averages (EMAs) of the features assigned to each row (see Section\,\ref{sssec:cast_tokenizer}). The overall tokenizer objective is given by:
\begin{equation}
\label{eq:tok_obj}
  \mathcal{L}_{\text{tok}}
    = \lambda_a\mathcal{L}_{\text{recons}} + \lambda_s\mathcal{L}_{\text{ctc}} + \lambda_{\text{int}}\mathcal{L}_{\text{inter-ctc}}
      + \beta\mathcal{L}_{\text{commit}},
\end{equation}
where $\lambda_a$, $\lambda_s$, $\lambda_{\text{int}}$, and $\beta$ are loss weighting coefficients.

\subsubsection{BiSAT-TTS Generation}
\label{sssec:tts_prob_form}
\textbf{BiSAT-TTS Text-Speech LM.} The text-speech LM autoregressively generates the codes. The text tokens and the speech codes are concatenated into a single sequence as input to the LM as follows:
\begin{equation}
\label{eq:clm_seq}
  \boldsymbol{x}_{\text{lm}}=\big[\langle\mathrm{SOT}\rangle;\,y_{1:U};\,\langle\mathrm{EOT}\rangle;\,z_{1:T};\,\langle\mathrm{EOS}\rangle\big]^{(\text{time})},
\end{equation}
where $\langle\mathrm{SOT}\rangle$ and $\langle\mathrm{EOT}\rangle$ mark the start and end of the text, and $\langle\mathrm{EOS}\rangle$ marks the end of speech. Let $z_{T+1}$ denote $\langle\mathrm{EOS}\rangle$.

The text-speech LM, parameterized by $\theta$, predicts a distribution over the $K$ codes and EOS at each step. We can factorize the conditional likelihood of the code sequence $\boldsymbol{z}_{1:T+1}$ given the text prompt $\boldsymbol{y}$ as:\looseness=-1
\begin{equation}
\label{eq:log-like}
  p_{\theta}(\boldsymbol{z}_{1:T+1}\mid\boldsymbol{y})
  =\prod_{t=1}^{T+1}p_{\theta}\big(z_t\mid\boldsymbol{z}_{<t},\boldsymbol{y}\big).
\end{equation}
The parameters $\theta$ can be optimized by minimizing the negative log-likelihood of the code sequence $\boldsymbol{z}_{1:T+1}$:
\begin{equation}
\label{eq:ar_loss}
  \mathcal{L}_{\text{lm}}(\theta)
  =-\sum_{t=1}^{T+1}\log p_{\theta}\big(z_t\mid\boldsymbol{z}_{<t},\boldsymbol{y}\big).
\end{equation}
At inference, the prefix $\boldsymbol{x}_{\text{pre}}$ consists of the prompt transcript $\tilde{\boldsymbol{y}}$, the target text $\boldsymbol{y}$, and the prompt codes $\tilde{\boldsymbol{z}}_{1:T'}$. The text-speech LM generates the target codes autoregressively from this prefix and stops at $\langle\mathrm{EOS}\rangle$:
\begin{equation}
\label{eq:clm_infer}
\begin{aligned}
  \boldsymbol{x}_{\text{pre}}&=\big[\langle\mathrm{SOT}\rangle;\,\tilde{\boldsymbol{y}};\,\boldsymbol{y};\,\langle\mathrm{EOT}\rangle;\,\tilde{\boldsymbol{z}}_{1:T'}\big]^{(\text{time})},\\
  z_t&\sim q_t,\qquad t=1,2,\dots,
\end{aligned}
\end{equation}
where $q_t$ is obtained from the LM distribution using the sampling procedure detailed in Section\,\ref{sssec:cast_neural_codec_lm}.

\textbf{BiSAT-TTS Acoustic Decoder.} The acoustic decoder models the conditional distribution of mel spectrograms given the speech codes using conditional flow matching (CFM) \cite{lipman2023flow}. We use linear interpolation paths between Gaussian noise $\boldsymbol{\epsilon}\sim\mathcal{N}(0,\mathbf{I})$ and the target mel spectrogram $\boldsymbol{s}$ \cite{liu2023flow},
\begin{equation}
\label{eq:noise_sample}
  \boldsymbol{n}_t=t\boldsymbol{s}+(1-t)\boldsymbol{\epsilon},\qquad t\in[0,1].
\end{equation}
Following \cite{du2024cosyvoice}, we sample training time steps as $t \sim \mathcal{U}(0,1)$. Taking the derivative of the interpolant with respect to $t$, the conditional velocity field $v(t)$ is obtained as:
\begin{equation}
\label{eq:flow_ode}
  v(t)=\frac{d\boldsymbol{n}_t}{dt}=\boldsymbol{s}-\boldsymbol{\epsilon}.
\end{equation}
A neural network parameterized by $\vartheta$ estimates the conditional mean velocity conditioned on $t$, $\boldsymbol{n}_t$, the speech codes $\boldsymbol{z}_{1:T}$, a prompt mel spectrogram $\tilde{\boldsymbol{s}} \in \mathbb{R}^{F \times T_{\text{prompt}}}$ extracted from reference speech, and a speaker condition $\boldsymbol{\xi}\in\mathbb{R}^{192}$ extracted by a frozen speaker encoder, such as CAM++ \cite{wang2023campplus}. We optimize the CFM by minimizing the expected squared error between the predicted and true velocities over the training distribution $\mathcal{D}$:\looseness=-1
\begin{equation}
\label{eq:flow_loss}
  \mathcal{L}_{\text{cfm}}(\vartheta)
  = \underset{\mathcal{D},\,t,\,\boldsymbol{\epsilon}}{\mathbb{E}}\;
    \Big\lVert v_{\vartheta}\big(t,\boldsymbol{n}_t\mid
      \boldsymbol{z}_{1:T},\tilde{\boldsymbol{s}},\boldsymbol{\xi}\big)
      -v(t)\Big\rVert_2^2,
\end{equation}
where $(\boldsymbol{s},\boldsymbol{z}_{1:T},\tilde{\boldsymbol{s}},\boldsymbol{\xi})\sim\mathcal{D}$.

At inference, we generate a mel spectrogram by solving the following ordinary differential equation (ODE) with a first-order Euler solver:\looseness=-1
\begin{equation}
\label{eq:solve_ode}
  \frac{d\boldsymbol{s}_t}{dt}=v_{\vartheta}(t,\boldsymbol{s}_t
  \mid\boldsymbol{z}_{1:T},\tilde{\boldsymbol{s}},\boldsymbol{\xi}),\qquad
  \boldsymbol{s}_0=\kappa\boldsymbol{\epsilon},
\end{equation}
from $t=0$ to $t=1$. $\kappa$ is a temperature that scales the noise. To solve Eq.\,(\ref{eq:solve_ode}), we discretize the time interval $[0,1]$ at time points $t_i$ obtained by applying a schedule to the uniform grid $u_i=i/J$ for $i=0,\ldots,J$. $J$ is the number of integration steps. We consider the cosine schedule $t_i=1-\cos(\pi u_i/2)$ \cite{du2024cosyvoice2,du2025cosyvoice3} and the PolyShift schedule $t_i=u_i^{\zeta}/\bigl(u_i^{\zeta}+\sigma(1-u_i^{\zeta})\bigr)$ \cite{chen2026wavtts}, where larger $\zeta$ and $\sigma$ place more steps toward $t=0$.

\subsection{Architecture}
\label{ssec:arch}

\subsubsection{BiSAT Tokenizer}
\label{sssec:cast_tokenizer}
\textbf{Dual Encoders.} The acoustic encoder $\mathcal{E}_a$ and semantic encoder $\mathcal{E}_s$ each stack 17 Conformer blocks (dim. 512, 8 attention heads) \cite{gulati2020conformer}. Conformer combines convolution for local acoustic structure with attention mechanisms for longer-range context. During training, we mask frequency bands and time patches in the semantic input using NeMo's spectral augmentation \cite{kuchaiev2019nemo,park2019specaugment} to enhance contextualization. The acoustic encoder receives unaugmented mel-spectrogram features. Let $\mathcal{A}$ denote the masking operation. The encoder outputs are given by:\looseness=-1
\begin{equation}
\label{eq:tok_enc_forward}
\begin{aligned}
    \boldsymbol{e}_a &= f_{\mathcal{E}_{\psi_a}}\bigl(\boldsymbol{m}\bigr), \\
    \boldsymbol{e}_s &= f_{\mathcal{E}_{\psi_s}}\bigl(\mathcal{A}(\boldsymbol{m})\bigr),
\end{aligned}
\end{equation}
with $\boldsymbol{e}_a \in \mathbb{R}^{D_a \times T}$ and $\boldsymbol{e}_s \in \mathbb{R}^{D_s \times T}$. The encoder outputs are then $\ell_2$-normalized and quantized.

\textbf{Shared-index Vector Quantizer (SVQ).} Motivated by the shared-index principle of TokenFlow \cite{qu2024tokenflow}, we design a shared-index quantization scheme using two codebooks of size $K \times D_{\text{codebook}}$, one for each branch, with $D_{\text{codebook}} = 512$. By Eq.\,(\ref{eq:shared_map}), the quantizer selects one shared index $z_t \in \{1,\dots,K\}$ per frame, so the frame-level discrete bottleneck remains $\log_2 K$ bits. This design has two key properties.

\textit{(i) Separate continuous representation spaces.}\enspace During training, each shared index retrieves paired acoustic and semantic entries $\boldsymbol{c}_{z_t,a}$ and $\boldsymbol{c}_{z_t,s}$, which are passed to their respective decoders. Learning the two continuous spaces and codebooks separately allows each branch to develop task-specific representations without requiring acoustic detail and linguistic units to be encoded in the same continuous feature vector. At the same time, the joint distance in Eq.\,(\ref{eq:shared_map}) allows frames that are similar in one branch but different in the other to be assigned to different codebook indices, partitioning the $K$ paired entries according to both acoustic and semantic structure. Together with spherical k-means initialization, EMA updates, and dead-code replacement, this yields training codebook utilization of 100\% for BiSAT-32k and 99.8\% for BiSAT-64k, as shown in Table\,\ref{tab:tokenizer_eval}.\looseness=-1

\textit{(ii) Bidirectional alignment.}\enspace The shared index $z_t$ in Eq.\,(\ref{eq:shared_map}) depends on both $\bar{\boldsymbol{e}}_{a,t}$ and $\bar{\boldsymbol{e}}_{s,t}$, grouping frames under the joint acoustic and semantic distance. Mel-spectrogram reconstruction updates the acoustic encoder and CTC updates the semantic encoder, while EMA updates in Eq.\,(\ref{eq:ema}) adapt both codebooks from the same assigned frames. Through shared-index assignment, updates in either encoder affect the quantization assignments of both branches, coupling their optimization without relying on fixed semantic teacher features \cite{zhang2024speechtokenizer,defossez2024moshi,jo2026lmspt,ye2025xcodec,ye2025llasa}. Downstream models such as the text-speech LM, acoustic decoder, and understanding probes receive only the discrete code sequence $\boldsymbol{z}_{1:T}$ from the tokenizer and learn their own embedding tables without accessing $\boldsymbol{c}_a$ or $\boldsymbol{c}_s$, yet they benefit directly from the acoustic and linguistic structure captured by the code stream, as evaluated in Section\,\ref{sec:exp}.\looseness=-1

\textbf{Codebook Initialization and Updates.} We sample stage-1 encoder features (see Section\,\ref{ssec:exp_setup}) from each training language and initialize the codebooks with spherical k-means under the joint distance in Eq.\,(\ref{eq:shared_map}). We update them using EMAs \cite{razavi2019generating} of the assigned features. Since both branches select the same index at each frame, they share an EMA count $C_k$ of frames assigned to index $k$. Each branch maintains its own EMA feature sum $\boldsymbol{S}_{k,b}$. For each index $k$ and branch $b\in\{a,s\}$, the codebook updates are given by:\looseness=-1
\begin{equation}
\label{eq:ema}
\begin{aligned}
  C_k &\leftarrow \eta C_k + (1-\eta)\, n_k, \\
  \boldsymbol{S}_{k,b} &\leftarrow \eta \boldsymbol{S}_{k,b} + (1-\eta) \sum_{t:\, z_t = k} \bar{\boldsymbol{e}}_{b,t}, \\
  \boldsymbol{c}_{k,b} &\leftarrow (\boldsymbol{S}_{k,b}/C_k)\,/\,\lVert \boldsymbol{S}_{k,b}/C_k \rVert_2, \quad (n_k > 0),
\end{aligned}
\end{equation}
where $n_k$ is the number of frames assigned to $k$ in the current batch and $\eta$ is the EMA decay factor. We also track each index's usage with an EMA of $n_k/\sum_{k'}n_{k'}$, i.e., the proportion of frames in the current batch assigned to index $k$. Every $\Delta_{\text{dead}}$ steps, a code pair whose EMA frequency is below $\nu_{\text{dead}}/K$ is treated as a dead code pair. We then replace it with paired encoder features drawn from a reservoir of recent frames and reset its embeddings and EMA statistics together.

\textbf{Dual Decoders.} The tokenizer decoders are used only during training, supervising the two branches with mel-spectrogram regression and CTC rather than adversarial waveform synthesis, while waveform generation is handled separately in TTS. At each frame, the shared index $z_t$ selects the acoustic and semantic entries $\hat{\boldsymbol{z}}_{a,t}=\boldsymbol{c}_{z_t,a}$ and $\hat{\boldsymbol{z}}_{s,t}=\boldsymbol{c}_{z_t,s}$. Let $\boldsymbol{q}_b$ denote the output of the rotation-trick parameterization in Eq.\,(\ref{eq:rotation_trick}) for $b\in\{a,s\}$. The acoustic decoder $\mathcal{D}_a$ stacks 9 ConvNeXt V2 blocks \cite{woo2023convnextv2} with $2\times$ convolutional temporal upsampling after blocks 4 and 7 to increase the frame rate from 25\,Hz to 100\,Hz, following \cite{siuzdak2024vocos,kim2024clamtts}. The semantic decoder $\mathcal{D}_s$ stacks 9 Conformer blocks ($512$-d, $8$ heads, $2048$-d feed-forward network (FFN)) followed by a CTC head. The two decoders produce:\looseness=-1
\begin{equation}
\label{eq:tok_dec_forward}
\begin{aligned}
  \hat{\boldsymbol{m}} &= f_{\mathcal{D}_{\gamma_a}}\big(\sqrt{D_a}\, \boldsymbol{q}_a\big), \\
  p_{\gamma_s}(\pi_t \mid \hat{\boldsymbol{z}}_{s;1:T})
    &= \left[\operatorname{softmax}\!\left(f_{\mathcal{D}_{\gamma_s}}(\sqrt{D_s}\,\boldsymbol{q}_s)\right)\right]_{t,\pi_t}.
\end{aligned}
\end{equation}

\textbf{Code Extraction.} Spectrogram augmentation is disabled when extracting codes. The frozen tokenizer uses both encoders and the same joint nearest-neighbor lookup to produce $\boldsymbol{z}_{1:T}$. These tokens serve as LM targets and acoustic-decoder inputs.\looseness=-1

\subsubsection{BiSAT-TTS Text-Speech LM}
\label{sssec:cast_neural_codec_lm}
We initialize the text embedding layer and Transformer backbone of the text-speech LM from MiniCPM4-0.5B \cite{minicpm2025minicpm4} and continue training on BiSAT codes using Eq.\,(\ref{eq:ar_loss}). As in MiniCPM4-0.5B, the backbone uses grouped-query attention, SwiGLU activations, and RMSNorm with LongRoPE positional encoding. The token $\langle\mathrm{SOT}\rangle$ reuses MiniCPM4's pretrained text beginning-of-sequence token. We randomly initialize an audio embedding table of $K+2$ entries for the $K$ speech codes, $\langle\mathrm{EOT}\rangle$, and $\langle\mathrm{EOS}\rangle$, together with a bias-free output head over $K+1$ classes for the $K$ speech codes and $\langle\mathrm{EOS}\rangle$.\looseness=-1

\textbf{Sampling During Decoding.} We use top-$k$ and top-$p$ sampling with a repetition penalty \cite{hu2026qwen3tts}. Let $\mathcal{H}_t=\{z_1,\ldots,z_{t-1}\}$ be the set of codes already generated for the target utterance, excluding reference codes. Let $o_t(v)$ denote the logit for candidate code $v$ at step $t$. For $v \in \mathcal{H}_t$, the repetition penalty is applied as follows:
\begin{equation}
\label{eq:ras}
  \tilde{o}_t(v)=
  \begin{cases}
    o_t(v)/\rho, & o_t(v)\ge0,\\
    \rho o_t(v), & o_t(v)<0,
  \end{cases}
\end{equation}
where $\rho\ge1$ is the repetition penalty. Other logits remain unchanged. We sample from the distribution over the top-$k$ and top-$p$ logits by scaling with a temperature $\tau$ and applying the softmax function. For $U$ target text tokens and $T'$ reference speech codes (see Eq.\,(\ref{eq:clm_infer})), let $L_{\max}$ and $L_{\min}$ denote maximum and minimum generation lengths, respectively. Inspired by \cite{du2024cosyvoice}, we set $L_{\max}=\min(L_{\text{cap}},\lfloor r_{\max}^{(\ell)}U\rfloor,C_{\text{audio}}-T')$ and $L_{\min}=\min(L_{\max},\max(0,\lfloor r_{\min}^{(\ell)}U\rfloor))$, where $C_{\text{audio}}>T'$ limits the total length of the reference and generated codes, $L_{\text{cap}}$ is generation cap length, and $r_{\min}^{(\ell)},r_{\max}^{(\ell)}$ are code-to-text ratios for language $\ell$. We mask EOS below $L_{\min}$ and stop at EOS or $L_{\max}$. Inference settings are given in Section\,\ref{sssec:inference_details}.\looseness=-1

\subsubsection{BiSAT-TTS Acoustic Decoder}
\label{sssec:cast_mel_decoder}
The TTS acoustic decoder maps discrete codes to mel spectrograms. The vocoder and mel feature settings are specified in Section\,\ref{ssec:acoustic_features}.

\textbf{Decoder Architecture.} The acoustic decoder consists of a learned code embedding layer ($512$-d) and a diffusion Transformer (DiT) velocity decoder \cite{peebles2023dit}. The DiT follows the backbone configuration of CosyVoice\,3's acoustic decoder ($1024$-d, 16 attention heads, and $2048$-d FFN) \cite{du2025cosyvoice3}. The code embeddings pass through a pre-lookahead block used in \cite{du2025cosyvoice3} and are upsampled from 25\,Hz to the 50\,Hz mel frame rate. The velocity decoder stacks 22 DiT blocks with rotary positional embeddings \cite{su2024roformer}. By Eq.\,(\ref{eq:fm_precond_forward}), the dimension of the concatenated input is $D_{\text{feat}}=D_{\text{code}}+2F+192=864$.

\textbf{Forward Pass.} Each code is mapped to a $D_{\text{code}}$-dimensional vector by a learned embedding table $\operatorname{Emb}_{\vartheta}$ with $K$ entries. The pre-lookahead block $\mathcal{P}_{\vartheta}$ and the upsampler $\operatorname{Up}(\cdot)$ then give the code condition $\boldsymbol{h}$ at $T_{\text{mel}}$ frames. With $\tilde{\boldsymbol{s}}_{\text{pad}}$ the prompt mel spectrogram zero-padded to $T_{\text{mel}}$ frames and $\boldsymbol{\xi}$ broadcast over frames, the input to the velocity decoder is given by:
\begin{equation}
\label{eq:fm_precond_forward}
\begin{aligned}
  \boldsymbol{u} &= \operatorname{Emb}_{\vartheta}(\boldsymbol{z}_{1:T})
    \in\mathbb{R}^{D_{\text{code}}\times T},\\
  \boldsymbol{h} &= \operatorname{Up}\big(f_{\mathcal{P}_{\vartheta}}(\boldsymbol{u})\big)
    \in\mathbb{R}^{D_{\text{code}}\times T_{\text{mel}}},\\
  \boldsymbol{x}_t &= [\boldsymbol{n}_t;\,\boldsymbol{h};\,\boldsymbol{\xi}\mathbf{1}^{\top};\,\tilde{\boldsymbol{s}}_{\text{pad}}]^{(\text{feat})}.
\end{aligned}
\end{equation}
At inference, the concatenated prompt and generated codes $[\tilde{\boldsymbol{z}}_{1:T'};\,\boldsymbol{z}_{1:T}]^{(\text{time})}$ replace $\boldsymbol{z}_{1:T}$, and the initial prompt frames are removed from the generated mel spectrogram before vocoding. The DiT $\mathcal{V}_{\vartheta}$ predicts the velocity as follows:
\begin{equation}
\label{eq:fm_unet_forward}
  v_{\vartheta}(t,\boldsymbol{n}_t\mid\boldsymbol{z}_{1:T},\tilde{\boldsymbol{s}},\boldsymbol{\xi})
    = f_{\mathcal{V}_{\vartheta}}(t,\boldsymbol{x}_t).
\end{equation}

\textbf{Time Conditioning.} A sinusoidal embedding of the time step $t$ is passed through a two-layer multilayer perceptron (MLP) with Swish activation and injected into every DiT block through AdaLN-Zero \cite{peebles2023dit}.\looseness=-1

\textbf{Mel-Spectrogram Prompt and Speaker Conditioning.} We condition the decoder on the mel prompt $\tilde{\boldsymbol{s}}$ for reference prosody, timbre, and style, together with the CAM++ speaker embedding $\boldsymbol{\xi}$ \cite{wang2023campplus}. During training, we crop a random prefix of the target utterance as $\tilde{\boldsymbol{s}}$, with its length drawn uniformly from $0$ to $30\%$ of the utterance, and drop this prompt with probability $p_{\text{prompt}}=0.5$ for regularization. Note that the CFM loss is computed over the full sequence including the prefix. At inference, both conditions are extracted from the reference speech.\looseness=-1

\textbf{Classifier-Free Guidance.} During training, we jointly drop $\boldsymbol{h}$, $\boldsymbol{\xi}$, and $\tilde{\boldsymbol{s}}$ with probability $p_{\text{cfg}}=0.2$, sampled independently of prompt dropout, to learn unconditional velocity prediction \cite{ho2021classifierfree}. At inference, classifier-free guidance (CFG) combines the conditional and unconditional velocity predictions $v_{\text{cond}}$ and $v_{\text{uncond}}$ with guidance scale $g$:
\begin{equation}
\label{eq:cfg_velocity}
  v_{\text{cfg}}=v_{\text{cond}}+g\bigl(v_{\text{cond}}-v_{\text{uncond}}\bigr).
\end{equation}

\textbf{Streaming Support.} Following \cite{du2024cosyvoice2,du2025cosyvoice3}, each training batch uses full bidirectional attention with probability $p_{\text{offline}}=0.5$ and chunk-causal attention with chunk size $M=50$ mel frames otherwise. A single checkpoint can therefore run either offline or in streaming mode, decoding each chunk of $M/2$ codes with a 3-code lookahead and passing the resulting mel frames to the causal vocoder.\looseness=-1

\section{Experiments}
\label{sec:exp}
\subsection{Experimental Setup}
\label{ssec:exp_setup}
\subsubsection{Training Data and Preprocessing}
We train the BiSAT tokenizer and BiSAT-TTS on approximately 122\,k hours of speech across 11 languages. Table\,\ref{tab:training_data} lists all training datasets and their durations grouped by utterance language.
For codebook initialization, we use the first-stage encoders to extract features from training utterances sampled by language and duration (6144 utterances per language, duration bins at 2, 5, 10, and 20\,s). The initialization pool contains approximately 1.3\,M frames. Spherical k-means under the joint distance of Eq.\,(\ref{eq:shared_map}) then runs on this pool for 12 rounds in chunks of 1024 frames, and empty clusters are reseeded from the frames with the largest quantization error.\looseness=-1

\subsubsection{Acoustic Features and Vocoder}
\label{ssec:acoustic_features}
The tokenizer processes 16\,kHz speech, using an 80-bin log-mel spectrogram at 100\,Hz with per-utterance, per-bin normalization as both input and reconstruction target. The encoders subsample this representation by a factor of four to produce 25\,Hz codes. For TTS, we use the pretrained causal HiFT vocoder of CosyVoice 3 \cite{du2025cosyvoice3}, whose mel configuration uses 24\,kHz speech, 80 mel bins, and a hop size of 480 samples, corresponding to a 50\,Hz frame rate.\looseness=-1

\begin{table}[t]
\centering
\caption{Multilingual training data.}
\vspace{-0.45em}
\label{tab:training_data}
\fontsize{7.4}{9.2}\selectfont
\renewcommand{\arraystretch}{1.04}
\setlength{\tabcolsep}{0pt}
\begin{tabular*}{\columnwidth}{@{\hspace{3pt}\extracolsep{\fill}}lcl@{\hspace{3pt}}}
\specialrule{.11em}{.08em}{.20em}
Language & Hours (k) & Sources \\
\specialrule{.05em}{.18em}{.18em}
English & 51.7 & Emilia~\cite{he2024emilia}, LibriTTS-R~\cite{koizumi23_interspeech}, GLOBE V2~\cite{wang24b_interspeech}, VCTK~\cite{yamagishi2019vctk} \\
Chinese & 42.6 & Emilia, AISHELL-3~\cite{shi2021aishell3} \\
German & 8.4 & Emilia, Emilia-YODAS~\cite{li2024yodas}, CML-TTS~\cite{oliveira2023cmltts}, HUI~\cite{puchtler2021hui} \\
French & 8.1 & Emilia, Emilia-YODAS, CML-TTS \\
Korean & 7.1 & Emilia, Emilia-YODAS \\
Japanese & 2.5 & Emilia, Emilia-YODAS, JVS~\cite{takamichi2019jvs} \\
Dutch & 0.756 & CML-TTS \\
Spanish & 0.436 & CML-TTS \\
Italian & 0.129 & CML-TTS \\
Portuguese & 0.077 & CML-TTS \\
Polish & 0.044 & CML-TTS \\
\specialrule{.05em}{.18em}{.18em}
Total & 122 & -- \\
\specialrule{.11em}{.20em}{.08em}
\end{tabular*}
\end{table}

\begin{table*}[t]
  \centering
  \caption{Speech tokenizer specifications and reconstruction on LibriSpeech-PC test-clean \cite{meister2023librispeechpc,chen2025f5tts} and Common Voice 27 \cite{ardila2020commonvoice}, with WER for EN, FR, and DE and CER for ZH, JA, and KO. BiSAT-32k/64k use $K=32768/65536$. Util.\ is over the training corpus for BiSAT and from original reports for baselines. $N_q$ denotes RVQ layers, and cosine/PolyShift denote acoustic-decoder ODE schedules. \textbf{Bold} and \underline{underline} denote best and second best.}
  \vspace{-0.52em}
  \label{tab:tokenizer_eval}
  \fontsize{8.8}{10.6}\selectfont
  \setlength{\tabcolsep}{0pt}
  \setlength{\cmidrulekern}{0.3em}
  \renewcommand{\arraystretch}{1.03}
  \begin{tabular*}{\textwidth}{@{\hspace{3pt}\extracolsep{\fill}}llcccccccccccccc@{\hspace{3pt}}}
    \specialrule{.11em}{.08em}{.20em}
    \multirow{2}{*}[-3pt]{Category} & \multirow{2}{*}[-3pt]{Tokenizer} & \multicolumn{5}{c}{Specifications} & \multicolumn{3}{c}{LibriSpeech-PC} & \multicolumn{6}{c}{Common Voice WER/CER$\downarrow$} \\
    \cmidrule(lr){3-7} \cmidrule(lr){8-10} \cmidrule(l){11-16}
    & & \makecell{Rate\\(Hz)} & \makecell{Bitrate\\(bps)} & \makecell{Codebook\\size} & \makecell{Util.\\(\%)} & \makecell{Data\\(h)} & WER$\downarrow$ & SIM$\uparrow$ & UTMOS$\uparrow$ & EN & FR & DE & ZH & JA & KO \\
    \specialrule{.05em}{.18em}{.18em}
    -- & Ground truth & -- & -- & -- & -- & -- & 1.89 & 1.000 & 3.21 & 11.01 & 13.27 & 6.35 & 9.93 & 15.06 & 6.51 \\
    \specialrule{.05em}{.18em}{.18em}
    \multirow{2}{*}{Acoustic}
      & FocalCodec~\cite{dellalibera2025focalcodec} & 25 & 325 & $8192$ & 99.8 & 585 & 2.94 & 0.960 & 2.94 & 25.52 & -- & -- & -- & -- & -- \\
      & WavTokenizer~\cite{ji2025wavtokenizer} & 40 & 480 & $4096$ & 100 & 80k & 9.22 & 0.933 & 3.00 & 47.04 & -- & -- & -- & -- & -- \\
    \specialrule{.05em}{.18em}{.18em}
    \multirow{2}{*}{Semantic}
      & CosyVoice 2~\cite{du2024cosyvoice2} & 25 & 317 & $6561$ & 100 & 200k & 3.21 & 0.968 & \textbf{3.51} & 24.96 & 31.57 & 20.61 & 15.02 & 25.09 & 15.34 \\
      & CosyVoice 3~\cite{du2025cosyvoice3} & 25 & 317 & $6561$ & 100 & 530k & 2.26 & 0.969 & 3.25 & 17.49 & 20.36 & 12.30 & 13.19 & \textbf{19.33} & 11.89 \\
    \specialrule{.05em}{.18em}{.18em}
    \multirow{4}{*}{Unified}
      & SpeechTokenizer ($N_q=1$)~\cite{zhang2024speechtokenizer} & 50 & 500 & $1024$ & 95.9 & 960 & 4.59 & 0.828 & 1.78 & 54.51 & -- & -- & -- & -- & -- \\
      & SpeechTokenizer ($N_q=2$) & 50 & 1000 & $1024$ & 95.9 & 960 & 3.65 & 0.905 & 1.54 & 46.40 & -- & -- & -- & -- & -- \\
      & LM-SPT ($N_q=2$)~\cite{jo2026lmspt} & 12.5 & 325 & \makecell{$16384/4096$} & \makecell{86/98} & 100k & 2.55 & 0.935 & 2.67 & 23.63 & 27.31 & 18.79 & 16.33 & 25.22 & 16.47 \\
      & X-Codec2~\cite{ye2025xcodec2release} & 50 & 800 & $65536$ & 99.0 & 150k & 2.27 & \textbf{0.981} & 3.23 & 19.13 & 20.74 & 13.10 & 13.72 & \underline{21.09} & 11.86 \\
    \specialrule{.05em}{.18em}{.18em}
    \multirow{4}{*}{Ours}
      & BiSAT-32k (cosine) & 25 & 375 & $32768$ & 100 & 122k & 2.18 & 0.976 & 3.26 & 16.49 & 19.35 & \underline{11.81} & 13.14 & 21.10 & 10.40 \\
      & BiSAT-32k (PolyShift) & 25 & 375 & $32768$ & 100 & 122k & \textbf{2.15} & 0.976 & 3.26 & 16.33 & 18.92 & 11.90 & \underline{12.93} & 21.25 & 10.34 \\
      & BiSAT-64k (cosine) & 25 & 400 & $65536$ & 99.8 & 122k & \underline{2.16} & \underline{0.977} & \underline{3.28} & \textbf{16.17} & \underline{18.79} & \textbf{11.76} & \textbf{12.80} & 21.47 & \textbf{10.15} \\
      & BiSAT-64k (PolyShift) & 25 & 400 & $65536$ & 99.8 & 122k & 2.18 & 0.976 & \underline{3.28} & \underline{16.22} & \textbf{18.62} & 11.91 & \textbf{12.80} & 21.42 & \underline{10.25} \\
    \specialrule{.11em}{.20em}{.08em}
  \end{tabular*}
\end{table*}

\begin{table*}[t]
  \centering
  \caption{Understanding evaluation on discrete codes. ASR is reported as WER for EN, FR, and DE and as CER for ZH and JA. ASR probes are evaluated zero-shot on Common Voice and FLEURS. PR is PER on LibriSpeech test-clean, ASV is EER on the VoxCeleb1 test set, SID is accuracy on VoxCeleb1, and LID is accuracy on the FLEURS test and VoxLingua107 development sets. ER is weighted accuracy averaged over five folds on IEMOCAP.}
  \vspace{-0.52em}
  \label{tab:tokenizer_understanding}
  \fontsize{8.8}{10.6}\selectfont
  \setlength{\tabcolsep}{0pt}
  \setlength{\cmidrulekern}{0.3em}
  \renewcommand{\arraystretch}{1.03}
  \begin{tabular*}{\textwidth}{@{\hspace{3pt}\extracolsep{\fill}}llccccccccccccccc@{\hspace{3pt}}}
    \specialrule{.11em}{.08em}{.20em}
    \multirow{3}{*}[-3pt]{Category} & \multirow{3}{*}[-3pt]{Tokenizer} & \multicolumn{10}{c}{ASR$\downarrow$} & \multirow{3}{*}[-3pt]{PR$\downarrow$} & \multirow{3}{*}[-3pt]{ASV$\downarrow$} & \multirow{3}{*}[-3pt]{SID$\uparrow$} & \multirow{3}{*}[-3pt]{ER$\uparrow$} & \multirow{3}{*}[-3pt]{LID$\uparrow$} \\
    \cmidrule(lr){3-12}
    & & \multicolumn{5}{c}{Common Voice} & \multicolumn{5}{c}{FLEURS} & & & & & \\
    \cmidrule(lr){3-7} \cmidrule(lr){8-12}
    & & EN & FR & DE & ZH & JA & EN & FR & DE & ZH & JA & & & & & \\
    \specialrule{.05em}{.18em}{.18em}
    \multirow{2}{*}{Acoustic}
      & FocalCodec~\cite{dellalibera2025focalcodec} & 86.71 & -- & -- & -- & -- & 85.96 & -- & -- & -- & -- & 8.33 & \underline{18.0} & \underline{31.3} & 47.5 & -- \\
      & WavTokenizer~\cite{ji2025wavtokenizer} & 110.48 & -- & -- & -- & -- & 99.63 & -- & -- & -- & -- & 30.62 & 28.2 & 10.1 & 46.9 & -- \\
    \specialrule{.05em}{.18em}{.18em}
    \multirow{2}{*}{Semantic}
      & CosyVoice 2~\cite{du2024cosyvoice2} & 81.71 & 67.07 & 63.27 & 39.23 & 46.17 & 55.74 & 66.43 & 64.59 & 32.04 & 29.66 & 7.71 & 19.4 & 24.8 & 48.6 & 82.0 \\
      & CosyVoice 3~\cite{du2025cosyvoice3} & \underline{74.17} & \underline{57.80} & \underline{54.34} & \underline{34.18} & \textbf{41.37} & \underline{47.17} & \underline{56.04} & \underline{52.33} & \underline{28.16} & \textbf{26.34} & \underline{5.68} & 18.6 & 24.7 & \textbf{54.7} & \underline{90.6} \\
    \specialrule{.05em}{.18em}{.18em}
    Unified
      & X-Codec2~\cite{ye2025xcodec2release} & 89.67 & 75.01 & 80.76 & 51.43 & 54.72 & 87.66 & 71.20 & 75.77 & 46.45 & 49.08 & 8.82 & 23.2 & 5.6 & 44.5 & 23.6 \\
    \specialrule{.05em}{.18em}{.18em}
    Ours & BiSAT-32k & \textbf{37.71} & \textbf{52.55} & \textbf{43.95} & \textbf{32.55} & \underline{42.29} & \textbf{24.42} & \textbf{49.21} & \textbf{39.54} & \textbf{26.68} & \underline{27.99} & \textbf{1.92} & \textbf{14.3} & \textbf{40.4} & \underline{48.8} & \textbf{95.5} \\
    \specialrule{.11em}{.20em}{.08em}
  \end{tabular*}
\end{table*}

\subsubsection{Training Details} We use eight NVIDIA RTX 6000 Ada GPUs for each component, including both tokenizer training stages, the text-speech LM, and the CFM acoustic decoder. After tokenizer training, we freeze the tokenizer and extract code sequences to train the LM and acoustic decoder.

\textbf{BiSAT tokenizer.} Both encoders are initialized from the English-pretrained 17-layer NeMo SSL model \texttt{ssl\_en\_conformer\_large} \cite{kuchaiev2019nemo}.\footnote{\rule{0pt}{8.6pt}\url{https://catalog.ngc.nvidia.com/orgs/nvidia/nemo/models/ssl_en_conformer_large}} The CTC heads use a 16\,k multilingual byte-pair vocabulary. Our default uses $K=32768$ (BiSAT-32k, 375\,bps), and we also train $K=65536$ (BiSAT-64k, 400\,bps) with the same training settings (Section\,\ref{sssec:codebook_ablation}). Training proceeds in two stages. First, the acoustic and semantic branches are trained separately without VQ for 100\,k steps each under their respective reconstruction or CTC losses. Second, we insert the SVQ initialized from stage-1 encoder features and train both branches jointly using Eq.\,(\ref{eq:tok_obj}) for 200\,k steps. We set $\lambda_a=1$, $\beta=0.5$, $\omega=1$ in Eq.\,(\ref{eq:shared_map}), and $\lambda_s=0.7$ and $\lambda_{\text{int}}=0.3$ following OWSM-CTC \cite{peng2024owsmctc}. The EMA decay factor is $\eta=0.99$, and dead-code replacement is applied every $\Delta_{\text{dead}}=100$ steps with threshold $\nu_{\text{dead}}=0.01$. Both stages use AdamW with 5\,k warmup steps, BF16 mixed precision, and a batch budget of 312.5\,s of speech per rank. After warmup, the learning rate is held constant at $10^{-4}$ in stage 1, whereas in stage 2 it is held at $7.5\times 10^{-5}$ until 100\,k steps and then linearly decayed to $7.5\times 10^{-6}$ over the remaining 100\,k steps. Spectrogram augmentation on the semantic branch applies two frequency masks of up to 27 bins and ten time masks of up to 5\% of the sequence length.\looseness=-1

\textbf{Text-speech LM.} Starting from MiniCPM4-0.5B, we fine-tune all parameters in BF16 mixed precision for 400\,k steps with a batch budget of 24\,k text and speech tokens per rank, using AdamW with a peak learning rate of $10^{-4}$, 10\,k linear warmup steps, and cosine decay to $5\times 10^{-6}$. We extend its text tokenizer with 2,916 single-character Chinese, Japanese, and Korean tokens. We retain MiniCPM4's input, residual-branch, and output scaling factors of $12$, $1.4/\sqrt{24}$, and $256/1024$.\looseness=-1

\textbf{Acoustic decoder.} We train for 1\,M steps in BF16 mixed precision with a batch budget of 375\,s of target speech per rank, using AdamW \cite{loshchilov2019decoupled} with a peak learning rate of $10^{-4}$, 20\,k linear warmup steps, and linear decay to $10^{-6}$.

\subsubsection{Inference Details}
\label{sssec:inference_details}
For the acoustic decoder, we use the Euler solver with $J=10$, $\kappa=1$, and CFG scale $g=0.7$ under either the cosine schedule or the PolyShift schedule with $\zeta=2$ and $\sigma=1$. All reconstruction and zero-shot TTS benchmarks use offline decoding for a direct comparison with offline baselines.
Based on inference hyperparameter sweeps for text-speech LM sampling, we set top-$p=1.0$, $k=25$, repetition penalty $\rho=1.05$, and temperature $\tau=0.8$ for BiSAT-TTS-32k, and top-$p=0.9$, $k=25$, $\rho=1.1$, and $\tau=0.9$ for BiSAT-TTS-64k. For length counting, we exclude $\langle\mathrm{SOT}\rangle$ and $\langle\mathrm{EOT}\rangle$. We set $r_{\min}^{(\ell)}=1$ for English and Mandarin, $r_{\max}^{(\ell)}=30$ for English and $25$ for Mandarin, $C_{\text{audio}}=2048$, and $L_{\text{cap}}=1536$. Across all runs, no BiSAT-TTS output is empty, and one BiSAT-TTS-32k utterance reaches $L_{\max}$.\looseness=-1

\textbf{Tokenizer Baselines.} We compare BiSAT with publicly available tokenizers from each category in Section\,\ref{ssec:work_tokenization} on LibriSpeech-PC test-clean \cite{meister2023librispeechpc} and Common Voice 27 \cite{ardila2020commonvoice}, using their official checkpoints without retraining. These baselines include the AOTs FocalCodec \cite{dellalibera2025focalcodec} and WavTokenizer \cite{ji2025wavtokenizer}, the SETs of CosyVoice 2 \cite{du2024cosyvoice2} and CosyVoice 3 \cite{du2025cosyvoice3}, and the UTs SpeechTokenizer \cite{zhang2024speechtokenizer}, LM-SPT \cite{jo2026lmspt}, and X-Codec2 \cite{ye2025xcodec2release}. Since SpeechTokenizer and LM-SPT rely on multiple RVQ streams for full reconstruction, we include them in the reconstruction comparison and restrict our single-stream understanding probes to single-codebook tokenizers. Following \cite{defossez2023encodec}, let $N_q$ denote the number of RVQ layers used. For SpeechTokenizer, we evaluate $N_q\in\{1,2\}$ as in X-Codec2 \cite{ye2025xcodec2release}, and for LM-SPT we use $N_q=2$ so that its bitrate is comparable to BiSAT's. Notably, the CosyVoice tokenizers are the most comparable baselines to BiSAT, as they also produce a single 25\,Hz code stream at a similar bitrate and reconstruct speech through a CFM acoustic decoder.
In particular, BiSAT uses the same DiT backbone configuration as CosyVoice 3 for its CFM acoustic decoder, while using $512$-d code and $192$-d speaker conditions rather than $80$-d projections.\looseness=-1

\textbf{TTS Baselines.} We compare BiSAT-TTS with zero-shot TTS systems that generate discrete speech tokens with an AR LM, using their official checkpoints and default inference settings. The baselines include CosyVoice 1, 2, and 3 \cite{du2024cosyvoice,du2024cosyvoice2,du2025cosyvoice3}, Spark-TTS \cite{wang2025sparktts}, VoiceStar \cite{peng2025voicestar}, and LLaSA-1B and LLaSA-3B \cite{ye2025llasa} built on X-Codec2. For VoiceStar, the required target duration is estimated from the speaking rate of the prompt as in \cite{peng2025voicestar}. Note that the acoustic decoders of the CosyVoice series and BiSAT-TTS condition on a reference mel prompt and speaker embedding, whereas Spark-TTS, VoiceStar, and LLaSA synthesize waveforms from the code sequence alone.\looseness=-1

\subsubsection{Evaluation Metrics}
\textbf{Tokenizer Evaluation.} Reconstruction and understanding assess the acoustic and semantic information retained in the codes.

\textbf{Reconstruction.} \emph{Perceived naturalness} is measured by the neural mean opinion score (MOS) predictor UTMOSv2 \cite{baba2024utmosv2}. \emph{Speaker similarity} (SIM) is the cosine similarity between speaker embeddings of the resynthesized and original speech, extracted with the WavLM-Base+ speaker verification model \cite{chen2022wavlm}. On LibriSpeech-PC, we measure \emph{intelligibility} by the word error rate (WER) of Whisper-large-v3 \cite{radford2023whisper}. CosyVoice 2, CosyVoice 3, and BiSAT reconstruct speech without a mel prompt, using codes and a speaker condition extracted from the target utterance. The other tokenizers decode from the codes alone. All systems reconstruct the same input utterance without a separate reference utterance. For a harder in-the-wild setting, we also evaluate on the crowdsourced Common Voice 27 \cite{ardila2020commonvoice} test sets of English, French, German, Mandarin, Japanese, and Korean, using the same reconstruction protocol. The English set excludes the clips and speakers used in Seed-TTS test-en. For Mandarin, we compute the character error rate (CER) with Paraformer-zh. For the other languages, we use Whisper-large-v3 with Whisper normalizers and report CER for Japanese and Korean and WER for English, French, and German.\looseness=-1

\textbf{Understanding.} We adapt the SUPERB \cite{yang2021superb} and ML-SUPERB \cite{shi2023mlsuperb} protocols to discrete codes in place of continuous SSL features. For each tokenizer, we train a probe on the frozen codes at the tokenizer's own frame rate. The probe consists of a learned code embedding table and an 8-layer, 512-dimensional Transformer encoder, followed by a CTC head for recognition tasks or attentive pooling with a classification head for the other tasks. All tokenizers use the same probe and training recipe for each task.\looseness=-1

The English tasks follow SUPERB. ASR in the other languages and language identification (LID) follow ML-SUPERB. For English ASR, the probe is trained on LibriSpeech train-clean-100 \cite{panayotov2015librispeech}. For the other languages, we follow the monolingual ASR setting of ML-SUPERB and train a separate probe for each language on 100 hours of a single corpus. These are 100-hour subsets of MLS \cite{pratap2020mls} for French and German, subset S of WenetSpeech \cite{zhang2022wenetspeech} for Mandarin, and the small subset of ReazonSpeech \cite{yin2023reazonspeech} for Japanese. We evaluate ASR only in the zero-shot out-of-domain setting, using the Common Voice \cite{ardila2020commonvoice} and FLEURS \cite{conneau2023fleurs} test sets for each language. These corpora are not used to train the ASR probes. We train the phoneme recognition (PR) probe on LibriSpeech train-clean-100 and test it on test-clean. For automatic speaker verification (ASV) and speaker identification (SID), we use VoxCeleb1 \cite{nagrani2017voxceleb}. For emotion recognition (ER), a shared seven-class head is trained on IEMOCAP \cite{busso2008iemocap}, CREMA-D \cite{cao2014cremad}, RAVDESS \cite{livingstone2018ravdess}, and the English portion of ESD \cite{zhou2021esd}. It is evaluated on IEMOCAP with SUPERB's five-fold leave-one-session-out protocol. Following the multilingual LID setting of ML-SUPERB, a single LID probe covers 12 languages (Mandarin, Cantonese, English, Japanese, Korean, French, German, Dutch, Italian, Polish, Portuguese, and Spanish). It is trained on FLEURS and VoxLingua107 \cite{valk2021voxlingua107} and tested on the pooled FLEURS test and VoxLingua107 development sets. ASR is reported as WER, or as CER for Mandarin and Japanese, PR as phone error rate (PER), ASV as equal error rate (EER), SID and LID as accuracy, and ER as weighted accuracy over the five folds.\looseness=-1

\textbf{TTS Evaluation.} We evaluate cross-sentence zero-shot TTS on LibriSpeech-PC test-clean \cite{meister2023librispeechpc}, following the VALL-E protocol \cite{wang2023neural} and the evaluation settings of \cite{chen2025f5tts}. The prompt is a different utterance from the same speaker, and the text-speech LM takes the annotated prompt transcript followed by the target text. We also evaluate the cross-sentence setting on Seed-TTS test-zh and test-en \cite{anastassiou2024seedtts}. On LibriSpeech-PC, we report the WER of Whisper-large-v3 with its text normalizer, speaker similarity against the original prompt (SIM-o) computed with the WavLM-Base+ speaker verification model \cite{chen2022wavlm}, UTMOSv2 for perceived naturalness, and NISQA \cite{mittag2021nisqa} for overall speech quality. On Seed-TTS, we report SIM-o using the official WavLM-Large ECAPA-TDNN model, with the WER of Whisper-large-v3 on test-en, the CER of Paraformer-zh \cite{gao2022paraformer} on test-zh, and UTMOSv2 and NISQA on both sets. We average utterance-level scores over three seeds. For WER and CER, we assign empty clips an error rate of 100\% and leave error rates for nonempty clips uncapped. For similarity and quality metrics, we average scores over scorable clips. The failure rate (\emph{Fail}) is the percentage of clips across all evaluated sets that are empty or have a WER or CER of at least 100\%. For test-en, we use the Whisper normalizer instead of that of the Seed-TTS toolkit for more accurate results, since it excludes differences that exist only in the written form but are pronounced identically, such as number formats.\looseness=-1

\subsection{Experimental Results}
\subsubsection{Tokenizer Evaluation}
\textbf{Reconstruction.} Table\,\ref{tab:tokenizer_eval} shows that BiSAT preserves both intelligibility and speaker fidelity at a low bitrate without adversarial waveform training. Among tokenizers at comparable bitrates, BiSAT achieves the strongest intelligibility and speaker similarity on LibriSpeech-PC while trailing only CosyVoice 2 in UTMOSv2. Despite the joint contribution of the discrete codes and the target speaker condition to SIM for BiSAT and the CosyVoice tokenizers, BiSAT-32k also leads on SID and ASV in Table\,\ref{tab:tokenizer_understanding}, which isolate the effect of the speaker condition and show that the codes themselves retain strong speaker information. Moreover, BiSAT's margin over CosyVoice 3 rises from 0.11\% on clean LibriSpeech-PC to 0.4-1.7\% across English, French, German, and Korean on the in-the-wild Common Voice 27 sets despite using less than a quarter of CosyVoice 3's training data, with Japanese as the only exception. These results indicate that BiSAT's bidirectional alignment produces codes that generalize better across varied accents, microphones, and background noise than purely semantic or feature-fusion tokenization.\looseness=-1

Among the codecs that resynthesize the waveform from the codes alone using adversarial training, WavTokenizer has a UTMOSv2 close to the ground truth but about three times the WER of FocalCodec. Natural-sounding output thus does not ensure that the content is preserved, as discussed in Section\,\ref{sssec:work_acoustic_tok}. X-Codec2 achieves reconstruction WER close to BiSAT, yet the probes trained on its codes perform substantially worse on speech and phoneme recognition in Table\,\ref{tab:tokenizer_understanding}, showing that low reconstruction error alone does not mean linguistic information is readily accessible from the codes.\looseness=-1

\textbf{Understanding.} The frozen-code probes in Table\,\ref{tab:tokenizer_understanding} evaluate linguistic content through ASR, PR, and LID, speaker information through SID and ASV, and paralinguistic information through ER. The CosyVoice tokenizers rank second and third on most content tasks but obtain lower SID accuracy than BiSAT and FocalCodec (24.8\% and 24.7\% vs.\ 40.4\% and 31.3\%) and higher ASV error. Speaker recognition relies on both spectral and prosodic cues \cite{kajarekar2003speaker}, so these results suggest that SETs trained only on understanding tasks retain less speaker-related acoustic detail in their codes than BiSAT. FocalCodec shows the opposite pattern, with higher SID accuracy than the SETs but more recognition errors.\looseness=-1

\begin{figure}[t]
  \centering
  \definecolor{bisatAco}{HTML}{A63F12}%
  \definecolor{bisatSem}{HTML}{164C96}%
  \definecolor{bisatAcoP}{HTML}{E08214}%
  \definecolor{bisatSemP}{HTML}{3895C8}%
  \definecolor{bisatK1}{HTML}{116653}%
  \definecolor{bisatK5}{HTML}{7A4988}%
  \definecolor{bisatInk}{HTML}{1A1A1A}%
  \definecolor{bisatMuted}{HTML}{555555}%
  \definecolor{bisatGrid}{HTML}{DFDFDF}%
  \newcommand{\figssrm}{\IfFileExists{OT1SourceSerifFour-TLF.fd}{\fontencoding{OT1}\fontfamily{SourceSerifFour-TLF}\fontseries{m}\fontshape{n}\selectfont}{\IfFileExists{OT1SourceSerifPro-TLF.fd}{\fontencoding{OT1}\fontfamily{SourceSerifPro-TLF}\fontseries{m}\fontshape{n}\selectfont}{}}}%
  \newcommand{\figsssb}{\IfFileExists{OT1SourceSerifFour-LF.fd}{\fontencoding{OT1}\fontfamily{SourceSerifFour-LF}\fontseries{sb}\fontshape{n}\selectfont}{\IfFileExists{OT1SourceSerifPro-LF.fd}{\fontencoding{OT1}\fontfamily{SourceSerifPro-LF}\fontseries{sb}\fontshape{n}\selectfont}{\bfseries}}}%
  \newcommand{\figsssub}[1]{\raisebox{-0.24ex}{\scalebox{0.74}{#1}}}%
  \newcommand{\figssDelta}{\IfFileExists{OT1SourceSerifFour-TLF.fd}{{\fontencoding{OT1}\fontfamily{SourceSerifFour-TLF}\fontseries{m}\fontshape{n}\selectfont\char1}}{$\Delta$}}%
  \newsavebox{\figssarrbox}\sbox{\figssarrbox}{\raisebox{0.34ex}{\tikz[baseline=0pt]\draw[-{Latex[length=0.88mm,width=0.68mm]},line width=0.38pt,bisatInk] (0,0) -- (2.05mm,0);}}%
  \newsavebox{\figssarrsmbox}\sbox{\figssarrsmbox}{\raisebox{0.25ex}{\tikz[baseline=0pt]\draw[-{Latex[length=0.64mm,width=0.50mm]},line width=0.31pt,bisatMuted] (0,0) -- (1.42mm,0);}}%
  \newsavebox{\figssnearrbox}\sbox{\figssnearrbox}{\raisebox{0.04ex}{\tikz[baseline=0pt]\draw[-{Latex[length=0.78mm,width=0.62mm]},line width=0.36pt,black!78] (0,0) -- (1.35mm,1.35mm);}}%
  \resizebox{\columnwidth}{!}{%
  \begin{tikzpicture}[x=1mm,y=1mm,font=\figssrm\scriptsize,text=bisatInk]
    \foreach \gx in {10.464,20.321,30.179,40.036}
      \draw[bisatGrid,line width=0.35pt] (\gx,0) -- (\gx,34.5);
    \foreach \gy in {2.464,12.321,22.179,32.036}
      \draw[bisatGrid,line width=0.35pt] (8.00,\gy) -- (42.50,\gy);

    \draw[color=black!82,line width=0.52pt] (8.00,0.0) rectangle (42.50,34.5);
    \foreach \tx in {10.464,20.321,30.179,40.036}
      \draw[color=black!82,line width=0.45pt] (\tx,0) -- (\tx,0.75) (\tx,34.5) -- (\tx,33.75);
    \foreach \tx/\tl in {10.464/50\,k,20.321/100\,k,30.179/150\,k,40.036/200\,k}
      \node[anchor=base,inner sep=0pt,font=\figssrm\fontsize{7.0}{8.0}\selectfont] at (\tx,-2.65) {\tl};
    \foreach \ty in {2.464,12.321,22.179,32.036}
      \draw[color=black!82,line width=0.45pt] (8.00,\ty) -- (8.80,\ty) (42.50,\ty) -- (41.70,\ty);
    \foreach \ty/\tl in {2.464/0.02,12.321/0.05,22.179/0.08,32.036/0.11}
      \node[anchor=east,inner sep=0pt,font=\figssrm\fontsize{7.0}{8.0}\selectfont] at (7.40,\ty) {\tl};
    \node[anchor=base,inner sep=0pt,font=\figssrm\fontsize{7.6}{8.6}\selectfont] at (25.25,-5.75) {Training steps};
    \node[rotate=90,anchor=south,inner sep=0pt,font=\figssrm\fontsize{7.6}{8.6}\selectfont] at (1.55,17.25) {Excess distortion \textit{r}\kern-0.07em\figsssub{\textit{b}}};
    \node[anchor=base,inner sep=0pt,font=\figsssb\fontsize{7.8}{8.8}\selectfont] at (25.25,35.90) {(a) Excess distortion};

    \draw[bisatSemP,line width=1.12pt]
      (10.464,25.320) -- (20.321,22.717) -- (30.179,21.899) -- (40.036,20.030);
    \draw[bisatAcoP,line width=1.12pt]
      (10.464,31.655) -- (20.321,26.266) -- (30.179,22.872) -- (40.036,20.309);
    \foreach \px/\py in {10.464/25.320,20.321/22.717,30.179/21.899,40.036/20.030}
      \fill[bisatSemP] (\px,\py) circle[radius=0.56mm];
    \foreach \px/\py in {10.464/31.655,20.321/26.266,30.179/22.872,40.036/20.309}
      \fill[bisatAcoP] (\px,\py) circle[radius=0.56mm];

    \draw[bisatSem,line width=1.12pt]
      (10.464,4.869) -- (20.321,4.416) -- (30.179,4.239) -- (40.036,3.792);
    \draw[bisatAco,line width=1.12pt]
      (10.464,8.185) -- (20.321,6.033) -- (30.179,4.465) -- (40.036,3.516);
    \foreach \px/\py in {10.464/4.869,20.321/4.416,30.179/4.239,40.036/3.792}
      \fill[bisatSem] (\px,\py) circle[radius=0.56mm];
    \foreach \px/\py in {10.464/8.185,20.321/6.033,30.179/4.465,40.036/3.516}
      \fill[bisatAco] (\px,\py) circle[radius=0.56mm];

    \node[anchor=east,text=bisatMuted,inner sep=0.3pt,font=\figssrm\fontsize{7.1}{8.1}\selectfont] at (13.45,39.95) {Mean:};
    \draw[bisatAco,line width=1.25pt] (13.85,39.95) -- (16.35,39.95);
    \fill[bisatAco] (15.10,39.95) circle[radius=0.52mm];
    \node[anchor=west,inner sep=0.3pt,font=\figssrm\fontsize{7.4}{8.4}\selectfont] at (16.55,39.95) {\textit{r}\kern-0.07em\figsssub{\textit{a}}};
    \draw[bisatSem,line width=1.25pt] (19.65,39.95) -- (22.15,39.95);
    \fill[bisatSem] (20.90,39.95) circle[radius=0.52mm];
    \node[anchor=west,inner sep=0.3pt,font=\figssrm\fontsize{7.4}{8.4}\selectfont] at (22.35,39.95) {\textit{r}\kern-0.07em\figsssub{\textit{s}}};
    \draw[black!28,line width=0.45pt] (25.55,38.80) -- (25.55,41.10);
    \node[anchor=east,text=bisatMuted,inner sep=0.3pt,font=\figssrm\fontsize{7.1}{8.1}\selectfont] at (30.35,39.95) {\textit{P}\kern-0.05em\figsssub{95}:};
    \draw[bisatAcoP,line width=1.25pt] (30.75,39.95) -- (33.25,39.95);
    \fill[bisatAcoP] (32.00,39.95) circle[radius=0.52mm];
    \node[anchor=west,inner sep=0.3pt,font=\figssrm\fontsize{7.4}{8.4}\selectfont] at (33.45,39.95) {\textit{r}\kern-0.07em\figsssub{\textit{a}}};
    \draw[bisatSemP,line width=1.25pt] (36.55,39.95) -- (39.05,39.95);
    \fill[bisatSemP] (37.80,39.95) circle[radius=0.52mm];
    \node[anchor=west,inner sep=0.3pt,font=\figssrm\fontsize{7.4}{8.4}\selectfont] at (39.25,39.95) {\textit{r}\kern-0.07em\figsssub{\textit{s}}};

    \foreach \gx in {58.464,68.321,78.179,88.036}
      \draw[bisatGrid,line width=0.35pt] (\gx,0) -- (\gx,34.5);
    \foreach \gy in {2.464,12.321,22.179,32.036}
      \draw[bisatGrid,line width=0.35pt] (56.00,\gy) -- (90.50,\gy);

    \draw[black!48,dash pattern=on 2.2pt off 1.3pt,line width=0.65pt] (56.000,0.000) -- (90.500,34.500);
    \node[rotate=45,text=bisatMuted,fill=white,inner sep=0.35pt,font=\figssrm\fontsize{5.2}{6.1}\selectfont] at (73.95,17.95)
      {\figssDelta(\textit{s}\,\usebox{\figssarrsmbox}\,\textit{a})\,{=}\,\figssDelta(\textit{a}\,\usebox{\figssarrsmbox}\,\textit{s})};
    \node[anchor=north east,text=black!78,fill=white,inner sep=0.4pt,font=\figssrm\fontsize{6.2}{7.2}\selectfont] at (86.0,33.3)
      {Null (1,\,1)\,\usebox{\figssnearrbox}};

    \draw[color=black!82,line width=0.52pt] (56.00,0.0) rectangle (90.50,34.5);
    \foreach \tx in {58.464,68.321,78.179,88.036}
      \draw[color=black!82,line width=0.45pt] (\tx,0) -- (\tx,0.75) (\tx,34.5) -- (\tx,33.75);
    \foreach \tx/\tl in {58.464/0.16,68.321/0.24,78.179/0.32,88.036/0.40}
      \node[anchor=base,inner sep=0pt,font=\figssrm\fontsize{7.0}{8.0}\selectfont] at (\tx,-2.65) {\tl};
    \foreach \ty in {2.464,12.321,22.179,32.036}
      \draw[color=black!82,line width=0.45pt] (56.00,\ty) -- (56.80,\ty) (90.50,\ty) -- (89.70,\ty);
    \foreach \ty/\tl in {2.464/0.16,12.321/0.24,22.179/0.32,32.036/0.40}
      \node[anchor=east,inner sep=0pt,font=\figssrm\fontsize{7.0}{8.0}\selectfont] at (55.35,\ty) {\tl};
    \node[anchor=base,inner sep=0pt,font=\figssrm\fontsize{7.6}{8.6}\selectfont] at (73.25,-5.75) {\figssDelta(\textit{a}\,\usebox{\figssarrbox}\,\textit{s})};
    \node[rotate=90,anchor=south,inner sep=0pt,font=\figssrm\fontsize{7.6}{8.6}\selectfont] at (49.45,17.25) {\figssDelta(\textit{s}\,\usebox{\figssarrbox}\,\textit{a})};
    \node[anchor=base,inner sep=0pt,font=\figsssb\fontsize{7.8}{8.8}\selectfont] at (73.25,35.90) {(b) Information imbalance};

    \draw[bisatK5,line width=1.12pt]
      (79.140,15.463) -- (82.750,15.488) -- (87.383,12.629) -- (88.664,11.582);
    \draw[-{Latex[length=1.40mm,width=0.98mm]},bisatK5,line width=0.58pt] (80.145,15.470) -- (81.745,15.481);
    \draw[-{Latex[length=1.40mm,width=0.98mm]},bisatK5,line width=0.58pt] (84.386,14.479) -- (85.747,13.638);
    \foreach \px/\py in {79.140/15.463,82.750/15.488,87.383/12.629,88.664/11.582}
      \fill[bisatK5] (\px,\py) circle[radius=0.56mm];

    \draw[bisatK1,line width=1.12pt]
      (72.708,7.639) -- (76.799,7.565) -- (81.530,5.113) -- (83.156,3.980);
    \draw[-{Latex[length=1.40mm,width=0.98mm]},bisatK1,line width=0.58pt] (73.954,7.616) -- (75.553,7.588);
    \draw[-{Latex[length=1.40mm,width=0.98mm]},bisatK1,line width=0.58pt] (78.454,6.707) -- (79.875,5.971);
    \foreach \px/\py in {72.708/7.639,76.799/7.565,81.530/5.113,83.156/3.980}
      \fill[bisatK1] (\px,\py) circle[radius=0.56mm];

    \node[anchor=east,text=bisatK1,inner sep=0.6pt,font=\figssrm\fontsize{6.2}{7.1}\selectfont] at (71.8,7.6) {50\,k};
    \node[anchor=south,text=bisatK1,inner sep=0.6pt,font=\figssrm\fontsize{6.2}{7.1}\selectfont] at (76.2,8.3) {100\,k};
    \node[anchor=north east,text=bisatK1,inner sep=0.6pt,font=\figssrm\fontsize{6.2}{7.1}\selectfont] at (80.9,4.6) {150\,k};
    \node[anchor=west,text=bisatK1,inner sep=0.6pt,font=\figssrm\fontsize{6.2}{7.1}\selectfont] at (84.0,4.0) {200\,k};

    \node[anchor=north,text=bisatK5,inner sep=0.6pt,font=\figssrm\fontsize{6.2}{7.1}\selectfont] at (79.7,14.6) {50\,k};
    \node[anchor=south,text=bisatK5,inner sep=0.6pt,font=\figssrm\fontsize{6.2}{7.1}\selectfont] at (83.0,16.2) {100\,k};
    \node[anchor=south,text=bisatK5,inner sep=0.5pt,font=\figssrm\fontsize{6.2}{7.1}\selectfont] at (87.8,13.4) {150\,k};
    \node[anchor=north east,text=bisatK5,inner sep=0.6pt,font=\figssrm\fontsize{6.2}{7.1}\selectfont] at (87.5,11.2) {200\,k};

    \draw[bisatK1,line width=1.30pt] (61.2,39.95) -- (65.2,39.95);
    \fill[bisatK1] (63.2,39.95) circle[radius=0.56mm];
    \node[anchor=west,inner sep=0.5pt,font=\figssrm\fontsize{7.7}{8.7}\selectfont] at (65.5,39.95) {\textit{m}\,=\,1};
    \draw[bisatK5,line width=1.30pt] (75.4,39.95) -- (79.4,39.95);
    \fill[bisatK5] (77.4,39.95) circle[radius=0.56mm];
    \node[anchor=west,inner sep=0.5pt,font=\figssrm\fontsize{7.7}{8.7}\selectfont] at (79.7,39.95) {\textit{m}\,=\,5};
  \end{tikzpicture}}%
  \vspace{-0.85em}
  \caption{Joint SVQ trajectories: (a) excess distortion $r_b$ averaged over Seed-TTS test-en and test-zh; (b) directional information imbalance.}
  \label{fig:codebook_diagnostics}
\end{figure}
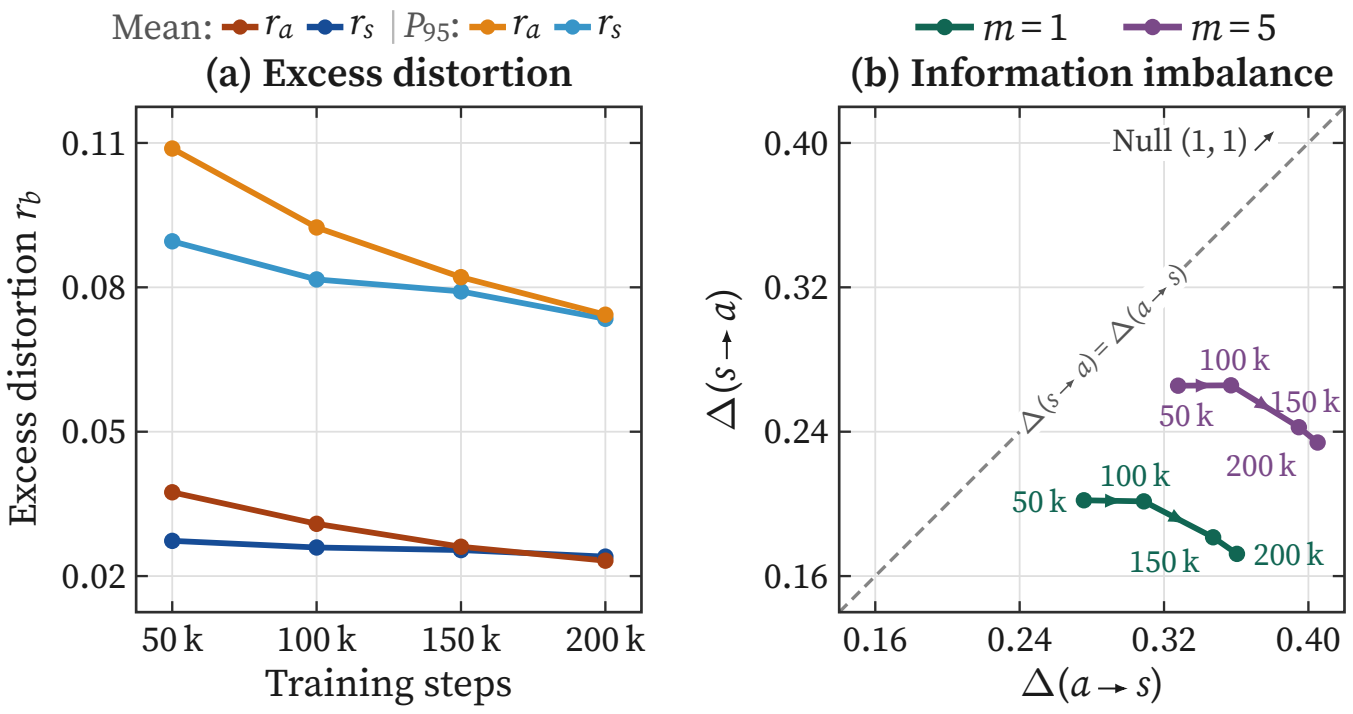

As shown by the reconstruction and probe results, the adversarially trained codecs have high recognition error rates on both test datasets, even for X-Codec2 with its semantic supervision, indicating that adversarial waveform reconstruction does not ensure code patterns that generalize across corpora. BiSAT outperforms X-Codec2 across all three task groups at less than half its bitrate. Unlike X-Codec2, which fuses acoustic and semantic features into one representation before quantization (see Section\,\ref{sssec:work_unified_tok}), BiSAT learns continuous representations in separate branches during training while selecting a shared codebook index at each frame to form a single code stream. Although the two systems also differ in encoder architecture and training supervision, so that this factor cannot be isolated from a cross-system comparison alone (see Section\,\ref{sssec:diag_analysis}), these results still show that BiSAT's code stream achieves strong representation quality within a compact single-stream framework. BiSAT also leads on all linguistic and speaker-related tasks except Japanese ASR, where CosyVoice 3 leads on both corpora, likely due to its much larger Japanese training data \cite{du2025cosyvoice3}. Together with the reconstruction results, these findings demonstrate that BiSAT preserves information useful for both linguistic understanding and acoustic reconstruction.\looseness=-1

The exception is ER, where CosyVoice 3 leads by a clear margin and BiSAT is on par with CosyVoice 2 and the AOTs. A likely reason is that the CosyVoice 3 tokenizer is trained with multi-task supervision including ER, whereas BiSAT receives no paralinguistic labels. Adding such supervision to the semantic branch is left for future work.\looseness=-1

As an additional implication, the strong ASR, PR, SID, and ASV probe results indicate that BiSAT codes could also support token-level reward modeling for reinforcement learning (RL) post-training (e.g., Token2Text content and multi-task rewards \cite{du2025cosyvoice3}).\looseness=-1

\subsubsection{Diagnostic Analysis}
\label{sssec:diag_analysis}
To analyze how BiSAT organizes its paired codebooks, we examine assignment alignment via excess distortion, codebook geometry via directional information imbalance, and branch specialization via TreeSHAP attribution.\looseness=-1

\textbf{Branch-wise Excess Quantization Distortion.} By Eq.\,(\ref{eq:shared_map}), both branches share index $z_t$ at each frame, so $z_t$ is not necessarily the nearest neighbor in each branch's own codebook. For branch $b\in\{a,s\}$ and frame $t$, let $d_{b,t}(k)=\lVert\bar{\boldsymbol{e}}_{b,t}-\boldsymbol{c}_{k,b}\rVert_2^2$ denote the squared Euclidean distance after $\ell_2$ normalization. Since nearest-neighbor lookup minimizes distortion for a fixed codebook \cite{gersho1992vector}, we measure the excess distortion from index sharing as $r_{b,t}=d_{b,t}(z_t)-\min_k d_{b,t}(k)\ge 0$, which vanishes only when $z_t$ is branch $b$'s own nearest entry. We evaluate $r_{b,t}$ ($\omega=1$) at 50\,k, 100\,k, 150\,k, and 200\,k stage-2 steps over Seed-TTS test-en ($1{,}088$) and test-zh ($2{,}020$), reporting the frame-weighted mean and 95th percentile ($P_{95}$) averaged across both sets. In Fig.\,\ref{fig:codebook_diagnostics}(a), both the mean and $P_{95}$ of $r_a$ and $r_s$ decrease steadily during training: the acoustic branch starts higher but drops faster, and both converge to nearly the same levels by 200\,k steps, indicating that joint training progressively aligns their quantization assignments.\looseness=-1

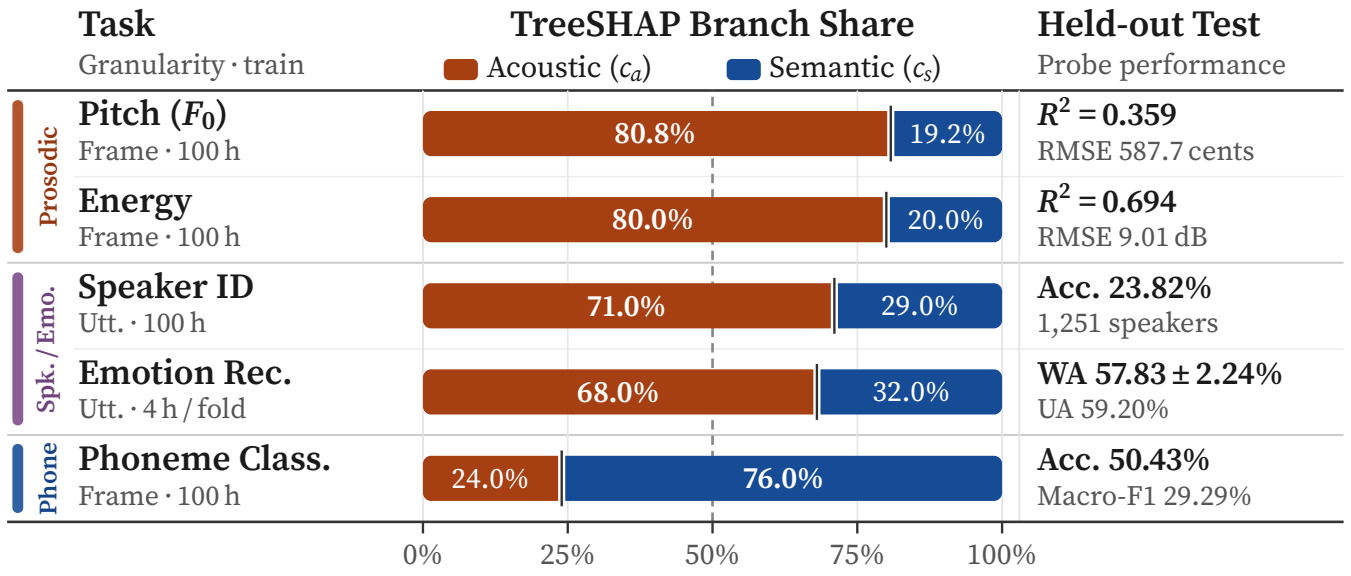
\begin{figure}[t]
  \centering
  \definecolor{bisatAco}{HTML}{A63F12}%
  \definecolor{bisatSem}{HTML}{164C96}%
  \definecolor{bisatK5}{HTML}{7A4988}%
  \definecolor{bisatInk}{HTML}{1A1A1A}%
  \definecolor{bisatMuted}{HTML}{555555}%
  \definecolor{bisatGrid}{HTML}{DFDFDF}%
  \newcommand{\shaprm}{\IfFileExists{OT1SourceSerifFour-TLF.fd}{\fontencoding{OT1}\fontfamily{SourceSerifFour-TLF}\fontseries{m}\fontshape{n}\selectfont}{\IfFileExists{OT1SourceSerifPro-TLF.fd}{\fontencoding{OT1}\fontfamily{SourceSerifPro-TLF}\fontseries{m}\fontshape{n}\selectfont}{}}}%
  \newcommand{\shapsb}{\IfFileExists{OT1SourceSerifFour-LF.fd}{\fontencoding{OT1}\fontfamily{SourceSerifFour-LF}\fontseries{sb}\fontshape{n}\selectfont}{\IfFileExists{OT1SourceSerifPro-LF.fd}{\fontencoding{OT1}\fontfamily{SourceSerifPro-LF}\fontseries{sb}\fontshape{n}\selectfont}{\bfseries}}}%
  \newcommand{\shapsub}[1]{\raisebox{-0.24ex}{\scalebox{0.74}{#1}}}%
  \resizebox{\columnwidth}{!}{%
  \begin{tikzpicture}[x=1mm,y=1mm,font=\shaprm\scriptsize,text=bisatInk]
    \draw[color=black!80,line width=0.50pt] (0.00,30.60) -- (90.00,30.60);

    \node[anchor=base west,inner sep=0pt,font=\shapsb\fontsize{7.1}{8.0}\selectfont] at (4.70,34.25) {Task};
    \node[anchor=base west,inner sep=0pt,text=bisatMuted,font=\shaprm\fontsize{5.5}{6.3}\selectfont] at (4.70,31.70) {Granularity\,\textperiodcentered\,train};

    \node[anchor=base,inner sep=0pt,font=\shapsb\fontsize{7.1}{8.0}\selectfont] at (47.50,34.25) {TreeSHAP Branch Share};
    \fill[bisatAco,rounded corners=0.35mm] (29.40,31.30) rectangle (31.70,32.72);
    \node[anchor=base west,inner sep=0pt,font=\shaprm\fontsize{5.8}{6.6}\selectfont] at (32.22,31.50) {Acoustic (\textit{c}\kern-0.05em\shapsub{\textit{a}})};
    \fill[bisatSem,rounded corners=0.35mm] (48.45,31.30) rectangle (50.75,32.72);
    \node[anchor=base west,inner sep=0pt,font=\shaprm\fontsize{5.8}{6.6}\selectfont] at (51.27,31.50) {Semantic (\textit{c}\kern-0.05em\shapsub{\textit{s}})};

    \node[anchor=base west,inner sep=0pt,font=\shapsb\fontsize{7.1}{8.0}\selectfont] at (69.30,34.25) {Held-out Test};
    \node[anchor=base west,inner sep=0pt,text=bisatMuted,font=\shaprm\fontsize{5.5}{6.3}\selectfont] at (69.30,31.70) {Probe performance};

    \draw[bisatGrid!55,line width=0.28pt] (4.50,24.80) -- (90.00,24.80);
    \draw[bisatGrid!92!black,line width=0.40pt] (0.00,19.00) -- (90.00,19.00);
    \draw[bisatGrid!55,line width=0.28pt] (4.50,13.20) -- (90.00,13.20);
    \draw[bisatGrid!92!black,line width=0.40pt] (0.00,7.40) -- (90.00,7.40);

    \draw[bisatGrid!80,line width=0.34pt] (68.15,1.55) -- (68.15,30.60);

    \draw[bisatGrid!85,line width=0.34pt] (28.00,1.55) -- (28.00,30.60);
    \draw[bisatGrid!75,line width=0.32pt] (37.75,1.55) -- (37.75,30.60);
    \draw[bisatGrid!75,line width=0.32pt] (57.25,1.55) -- (57.25,30.60);
    \draw[bisatGrid!85,line width=0.34pt] (67.00,1.55) -- (67.00,30.60);
    \draw[black!48,dash pattern=on 1.8pt off 1.15pt,line width=0.50pt] (47.50,1.55) -- (47.50,30.60);

    \fill[bisatAco!88,rounded corners=0.35mm] (0.35,19.65) rectangle (1.08,29.95);
    \node[rotate=90,anchor=center,inner sep=0pt,text=bisatAco!88!black,font=\shapsb\fontsize{4.9}{5.7}\selectfont] at (2.72,24.80) {Prosodic};

    \fill[bisatK5!88,rounded corners=0.35mm] (0.35,8.05) rectangle (1.08,18.35);
    \node[rotate=90,anchor=center,inner sep=0pt,text=bisatK5!90!black,font=\shapsb\fontsize{4.9}{5.7}\selectfont] at (2.72,13.20) {Spk.\,/\,Emo.};

    \fill[bisatSem!88,rounded corners=0.35mm] (0.35,2.15) rectangle (1.08,6.80);
    \node[rotate=90,anchor=center,inner sep=0pt,text=bisatSem!90!black,font=\shapsb\fontsize{4.8}{5.6}\selectfont] at (2.72,4.48) {Phone};

    \node[anchor=base west,inner sep=0pt,font=\shapsb\fontsize{6.8}{7.6}\selectfont] at (4.70,28.25) {Pitch (\textit{F}\kern-0.04em\shapsub{0})};
    \node[anchor=base west,inner sep=0pt,text=bisatMuted,font=\shaprm\fontsize{5.4}{6.1}\selectfont] at (4.70,25.92) {Frame\,\textperiodcentered\,100\,h};
    \fill[bisatAco] (59.264,26.18) [rounded corners=0.54mm] -- (28.00,26.18) -- (28.00,29.22) [sharp corners] -- (59.264,29.22) -- cycle;
    \fill[bisatSem] (59.744,29.22) [rounded corners=0.54mm] -- (67.00,29.22) -- (67.00,26.18) [sharp corners] -- (59.744,26.18) -- cycle;
    \draw[black!82,line width=0.45pt] (59.504,25.82) -- (59.504,29.58);
    \node[anchor=center,text=white,inner sep=0pt,font=\shapsb\fontsize{5.9}{6.7}\selectfont] at (43.63,27.70) {80.8\%};
    \node[anchor=center,text=white,inner sep=0pt,font=\shaprm\fontsize{5.5}{6.3}\selectfont] at (63.37,27.70) {19.2\%};
    \node[anchor=base west,inner sep=0pt,font=\shapsb\fontsize{6.3}{7.1}\selectfont] at (69.30,28.25) {\textit{R}\textsuperscript{2}\,{=}\,0.359};
    \node[anchor=base west,inner sep=0pt,text=bisatMuted,font=\shaprm\fontsize{5.4}{6.1}\selectfont] at (69.30,25.92) {RMSE 587.7\,cents};

    \node[anchor=base west,inner sep=0pt,font=\shapsb\fontsize{6.8}{7.6}\selectfont] at (4.70,22.45) {Energy};
    \node[anchor=base west,inner sep=0pt,text=bisatMuted,font=\shaprm\fontsize{5.4}{6.1}\selectfont] at (4.70,20.12) {Frame\,\textperiodcentered\,100\,h};
    \fill[bisatAco] (58.964,20.38) [rounded corners=0.54mm] -- (28.00,20.38) -- (28.00,23.42) [sharp corners] -- (58.964,23.42) -- cycle;
    \fill[bisatSem] (59.444,23.42) [rounded corners=0.54mm] -- (67.00,23.42) -- (67.00,20.38) [sharp corners] -- (59.444,20.38) -- cycle;
    \draw[black!82,line width=0.45pt] (59.204,20.02) -- (59.204,23.78);
    \node[anchor=center,text=white,inner sep=0pt,font=\shapsb\fontsize{5.9}{6.7}\selectfont] at (43.48,21.90) {80.0\%};
    \node[anchor=center,text=white,inner sep=0pt,font=\shaprm\fontsize{5.5}{6.3}\selectfont] at (63.22,21.90) {20.0\%};
    \node[anchor=base west,inner sep=0pt,font=\shapsb\fontsize{6.3}{7.1}\selectfont] at (69.30,22.45) {\textit{R}\textsuperscript{2}\,{=}\,0.694};
    \node[anchor=base west,inner sep=0pt,text=bisatMuted,font=\shaprm\fontsize{5.4}{6.1}\selectfont] at (69.30,20.12) {RMSE 9.01\,dB};

    \node[anchor=base west,inner sep=0pt,font=\shapsb\fontsize{6.8}{7.6}\selectfont] at (4.70,16.65) {Speaker ID};
    \node[anchor=base west,inner sep=0pt,text=bisatMuted,font=\shaprm\fontsize{5.4}{6.1}\selectfont] at (4.70,14.32) {Utt.\,\textperiodcentered\,100\,h};
    \fill[bisatAco] (55.466,14.58) [rounded corners=0.54mm] -- (28.00,14.58) -- (28.00,17.62) [sharp corners] -- (55.466,17.62) -- cycle;
    \fill[bisatSem] (55.946,17.62) [rounded corners=0.54mm] -- (67.00,17.62) -- (67.00,14.58) [sharp corners] -- (55.946,14.58) -- cycle;
    \draw[black!82,line width=0.45pt] (55.706,14.22) -- (55.706,17.98);
    \node[anchor=center,text=white,inner sep=0pt,font=\shapsb\fontsize{5.9}{6.7}\selectfont] at (41.73,16.10) {71.0\%};
    \node[anchor=center,text=white,inner sep=0pt,font=\shaprm\fontsize{5.6}{6.4}\selectfont] at (61.47,16.10) {29.0\%};
    \node[anchor=base west,inner sep=0pt,font=\shapsb\fontsize{6.3}{7.1}\selectfont] at (69.30,16.65) {Acc.\ 23.82\%};
    \node[anchor=base west,inner sep=0pt,text=bisatMuted,font=\shaprm\fontsize{5.4}{6.1}\selectfont] at (69.30,14.32) {1{,}251 speakers};

    \node[anchor=base west,inner sep=0pt,font=\shapsb\fontsize{6.8}{7.6}\selectfont] at (4.70,10.85) {Emotion Rec.};
    \node[anchor=base west,inner sep=0pt,text=bisatMuted,font=\shaprm\fontsize{5.4}{6.1}\selectfont] at (4.70,8.52) {Utt.\,\textperiodcentered\,4\,h\,/\,fold};
    \fill[bisatAco] (54.290,8.78) [rounded corners=0.54mm] -- (28.00,8.78) -- (28.00,11.82) [sharp corners] -- (54.290,11.82) -- cycle;
    \fill[bisatSem] (54.770,11.82) [rounded corners=0.54mm] -- (67.00,11.82) -- (67.00,8.78) [sharp corners] -- (54.770,8.78) -- cycle;
    \draw[black!82,line width=0.45pt] (54.530,8.42) -- (54.530,12.18);
    \node[anchor=center,text=white,inner sep=0pt,font=\shapsb\fontsize{5.9}{6.7}\selectfont] at (41.15,10.30) {68.0\%};
    \node[anchor=center,text=white,inner sep=0pt,font=\shaprm\fontsize{5.6}{6.4}\selectfont] at (60.89,10.30) {32.0\%};
    \node[anchor=base west,inner sep=0pt,font=\shapsb\fontsize{6.3}{7.1}\selectfont] at (69.30,10.85) {WA 57.83\,{\textpm}\,2.24\%};
    \node[anchor=base west,inner sep=0pt,text=bisatMuted,font=\shaprm\fontsize{5.4}{6.1}\selectfont] at (69.30,8.52) {UA 59.20\%};

    \node[anchor=base west,inner sep=0pt,font=\shapsb\fontsize{6.8}{7.6}\selectfont] at (4.70,5.03) {Phoneme Class.};
    \node[anchor=base west,inner sep=0pt,text=bisatMuted,font=\shaprm\fontsize{5.4}{6.1}\selectfont] at (4.70,2.70) {Frame\,\textperiodcentered\,100\,h};
    \fill[bisatAco] (37.103,2.96) [rounded corners=0.54mm] -- (28.00,2.96) -- (28.00,6.00) [sharp corners] -- (37.103,6.00) -- cycle;
    \fill[bisatSem] (37.583,6.00) [rounded corners=0.54mm] -- (67.00,6.00) -- (67.00,2.96) [sharp corners] -- (37.583,2.96) -- cycle;
    \draw[black!82,line width=0.45pt] (37.343,2.60) -- (37.343,6.36);
    \node[anchor=center,text=white,inner sep=0pt,font=\shaprm\fontsize{5.6}{6.4}\selectfont] at (32.55,4.48) {24.0\%};
    \node[anchor=center,text=white,inner sep=0pt,font=\shapsb\fontsize{5.9}{6.7}\selectfont] at (52.29,4.48) {76.0\%};
    \node[anchor=base west,inner sep=0pt,font=\shapsb\fontsize{6.3}{7.1}\selectfont] at (69.30,5.03) {Acc.\ 50.43\%};
    \node[anchor=base west,inner sep=0pt,text=bisatMuted,font=\shaprm\fontsize{5.4}{6.1}\selectfont] at (69.30,2.70) {Macro-F1 29.29\%};

    \draw[color=black!80,line width=0.50pt] (0.00,1.55) -- (90.00,1.55);
    \foreach \tx/\tl in {28.00/0\%,37.75/25\%,47.50/50\%,57.25/75\%,67.00/100\%} {
      \draw[color=black!80,line width=0.45pt] (\tx,1.55) -- (\tx,0.77);
      \node[anchor=base,inner sep=0pt,text=bisatMuted,font=\shaprm\fontsize{5.7}{6.5}\selectfont] at (\tx,-1.28) {\tl};
    }
  \end{tikzpicture}}%
  \vspace{-0.45em}
  \caption{Task-wise TreeSHAP attribution shares between the paired codebooks $\boldsymbol{c}_a$ and $\boldsymbol{c}_s$ alongside held-out XGBoost probe performance.}
  \label{fig:branch_shap}
\end{figure}

\textbf{Directional Information Imbalance.} Excess distortion reflects the frame-level cost of sharing $z_t$, rather than the relative geometry of $\boldsymbol{c}_a$ and $\boldsymbol{c}_s$. To measure whether neighboring indices in one codebook remain neighbors in the other, we compute directional information imbalance \cite{glielmo2022ranking}. Let $\mathcal{I}\subset\{1,\dots,K\}$ be a fixed subset of $V=|\mathcal{I}|$ active codebook indices, with squared Euclidean distances computed after $\ell_2$ normalization in each codebook. For each query index $i\in\mathcal{I}$, let $\mathcal{N}_m^{(a)}(i)\subset\mathcal{I}\setminus\{i\}$ denote its $m$ nearest neighbor indices in $\boldsymbol{c}_a$, and let $R_s(i,j)\in\{1,\dots,V-1\}$ be the distance rank of index $j$ from $i$ in $\boldsymbol{c}_s$. We define the information imbalance from $\boldsymbol{c}_a$ to $\boldsymbol{c}_s$ as:\looseness=-1
\begin{equation}
\label{eq:info_imbalance}
  \Delta(a\to s)=\frac{2}{V}\cdot\frac{1}{Vm}\sum_{i\in\mathcal{I}}\sum_{j\in\mathcal{N}_m^{(a)}(i)} R_s(i,j),
\end{equation}
and $\Delta(s\to a)$ is defined symmetrically by swapping $a$ and $s$. Under random index pairing, the expected rank is $V/2$, so $2/V$ normalizes the random baseline to $\mathbb{E}[\Delta]=1$, whereas smaller values mean $m$-nearest neighbors in the source codebook remain close in the target codebook. At the same four checkpoints, we evaluate $m\in\{1,5\}$ on the $V=28{,}173$ most frequent indices ($95\%$ of pooled training EMA usage) against $199$ random permutations. In Fig.\,\ref{fig:codebook_diagnostics}(b), all checkpoints lie well below $(1,1)$ ($p=0.005$). For $m=1$, $\Delta(s\to a)$ drops from $0.20$ to $0.17$ while $\Delta(a\to s)$ rises from $0.28$ to $0.36$, with smaller shifts after 150\,k steps (similarly for $m=5$). Neighbors in $\boldsymbol{c}_s$ thus map to closer ranks in $\boldsymbol{c}_a$, whereas the reverse weakens, plausibly because similar acoustic frames occur in different phonetic contexts. Thus, the two codebook spaces maintain non-random neighborhood correspondence while evolving asymmetrically. Because directional imbalance captures geometric correspondence rather than causal information transfer or gradient influence, these trajectories are consistent with coupled representation learning.\looseness=-1

\begin{table*}[t]
  \centering
  \caption{Cross-sentence zero-shot TTS results on LibriSpeech-PC and Seed-TTS over three seeds. UTMOS is UTMOSv2, NISQA is overall quality, Fail is the share of empty or $\ge 100\%$ WER/CER clips, Multi.\ is multiple datasets, and -32k/-64k denote $K=32768/65536$.}
  \vspace{-0.52em}
  \label{tab:tts_results}
  \fontsize{7.8}{9.1}\selectfont
  \setlength{\tabcolsep}{0pt}
  \setlength{\cmidrulekern}{0.3em}
  \renewcommand{\arraystretch}{1.05}
  \begin{tabular*}{\textwidth}{@{\hspace{1.5pt}\extracolsep{\fill}}lccccccccccccccc@{\hspace{1.5pt}}}
    \specialrule{.11em}{.08em}{.20em}
    \multirow{2}{*}[-2pt]{System} & \multirow{2}{*}[-2pt]{\makecell{Params\\(M)}} & \multirow{2}{*}[-2pt]{\makecell{Data\\(hours)}} & \multicolumn{4}{c}{LibriSpeech-PC} & \multicolumn{4}{c}{Seed-TTS (test-en)} & \multicolumn{4}{c}{Seed-TTS (test-zh)} & \multirow{2}{*}[-2pt]{\makecell{Fail$\downarrow$\\(\%)}} \\
    \cmidrule(lr){4-7} \cmidrule(lr){8-11} \cmidrule(lr){12-15}
    & & & WER$\downarrow$ & SIM-o$\uparrow$ & UTMOS$\uparrow$ & NISQA$\uparrow$ & WER$\downarrow$ & SIM-o$\uparrow$ & UTMOS$\uparrow$ & NISQA$\uparrow$ & CER$\downarrow$ & SIM-o$\uparrow$ & UTMOS$\uparrow$ & NISQA$\uparrow$ & \\
    \specialrule{.05em}{.18em}{.18em}
    Ground truth & -- & -- & 1.89 & 0.944 & 3.21 & 4.07 & 1.47 & 0.734 & 2.75 & 3.63 & 1.43 & 0.755 & 2.94 & 4.08 & -- \\
    \specialrule{.05em}{.18em}{.18em}
    CosyVoice 1~\cite{du2024cosyvoice} & 300 & 172k In-house & 2.87 & 0.906 & 3.25 & 4.26 & 3.85 & 0.512 & 3.10 & 4.13 & 3.59 & 0.643 & 3.00 & \textbf{4.35} & 0.157 \\
    CosyVoice 2~\cite{du2024cosyvoice2} & 500 & 167k In-house & 1.78 & 0.947 & \textbf{3.58} & \textbf{4.40} & 2.12 & 0.659 & \textbf{3.26} & 4.28 & 1.61 & \underline{0.752} & \textbf{3.10} & \underline{4.34} & 0.039 \\
    CosyVoice 3~\cite{du2025cosyvoice3} & 500 & 1M In-house & 1.72 & \textbf{0.951} & 3.43 & 4.28 & 1.55 & \textbf{0.698} & 3.07 & 4.15 & 1.43 & \textbf{0.779} & 2.89 & 4.22 & 0.063 \\
    Spark-TTS~\cite{wang2025sparktts} & 500 & 100k VoxBox & 2.69 & 0.938 & \underline{3.57} & 4.07 & 2.61 & 0.571 & \underline{3.16} & 4.07 & 2.70 & 0.658 & \underline{3.05} & 4.09 & 0.756 \\
    VoiceStar~\cite{peng2025voicestar} & 840 & 65k Multi. & 1.94 & 0.944 & 3.26 & 4.34 & 2.28 & 0.606 & 2.92 & 4.28 & -- & -- & -- & -- & 0.166 \\
    LLaSA-1B~\cite{ye2025llasa} & 1000 & 250k Multi. & 2.00 & 0.930 & 3.44 & 4.23 & 1.91 & 0.588 & 3.14 & 4.10 & 2.27 & 0.675 & 2.99 & 4.13 & 0.409 \\
    LLaSA-3B & 3000 & 250k Multi. & 2.17 & 0.931 & 3.43 & 4.22 & 2.38 & 0.594 & 3.11 & 4.12 & 1.69 & 0.682 & 2.96 & 4.12 & 0.512 \\
    \specialrule{.05em}{.18em}{.18em}
    BiSAT-TTS-32k (cosine) & 500 & 122k Multi. & 1.56 & \underline{0.950} & 3.43 & 4.37 & 1.30 & 0.669 & 3.03 & 4.27 & 1.39 & 0.734 & 2.89 & 4.28 & 0.031 \\
    BiSAT-TTS-32k (PolyShift) & 500 & 122k Multi. & \textbf{1.50} & 0.950 & 3.42 & 4.38 & 1.31 & \underline{0.669} & 3.04 & \underline{4.29} & 1.39 & 0.735 & 2.90 & 4.29 & \underline{0.024} \\
    BiSAT-TTS-64k (cosine) & 500 & 122k Multi. & 1.52 & 0.949 & 3.43 & 4.39 & \underline{1.23} & 0.667 & 3.05 & 4.29 & \underline{1.29} & 0.733 & 2.93 & 4.30 & \textbf{0.016} \\
    BiSAT-TTS-64k (PolyShift) & 500 & 122k Multi. & \underline{1.51} & 0.949 & 3.43 & \underline{4.39} & \textbf{1.21} & 0.667 & 3.06 & \textbf{4.30} & \textbf{1.27} & 0.734 & 2.92 & 4.30 & \textbf{0.016} \\
    \specialrule{.11em}{.20em}{.08em}
  \end{tabular*}
\end{table*}

\textbf{Task-wise Codebook Attribution.} To measure each codebook's contribution across speech properties, we train XGBoost probes \cite{chen2016xgboost} on the frozen continuous codebook vectors and compute branch-level TreeSHAP attributions \cite{lundberg2020local} (distinct from Table\,\ref{tab:tokenizer_understanding}'s sequence probes on discrete IDs). Frame-level probes take $[\boldsymbol{c}_{z_t,a};\,\boldsymbol{c}_{z_t,s}]\in\mathbb{R}^{1024}$ directly, while utterance-level probes mean-pool each $512$-d branch before concatenation. Per test sample, we sum the $512$ TreeSHAP values per branch, take absolute values, and normalize dataset averages into branch shares. Pitch ($F_0$), energy, and phoneme probes are trained on LibriSpeech train-clean-100 and evaluated on test-clean, using voiced $F_0$ from WORLD\footnote{\url{https://github.com/mmorise/World}} and phoneme alignments from the Montreal Forced Aligner.\footnote{\url{https://montreal-forced-aligner.readthedocs.io}} We use VoxCeleb1 for SID and five-fold cross-validation on IEMOCAP for four-class ER. Since both branches share $z_t$, these shares reflect probe reliance on $\boldsymbol{c}_a$ and $\boldsymbol{c}_s$ rather than prediction accuracy. In Fig.\,\ref{fig:branch_shap}, probes rely mainly on $\boldsymbol{c}_a$ (about $80\%$) for both $F_0$ and energy despite different $R^2$ scores ($0.359$ vs.\ $0.694$) and on $\boldsymbol{c}_s$ ($76\%$) for phonemes, whereas speaker and emotion probes draw on both codebooks.\looseness=-1

\subsubsection{Zero-shot TTS}
\textbf{Intelligibility.} As shown in Table\,\ref{tab:tts_results}, BiSAT-TTS achieves lower WER, CER, and Fail rates than all baselines, far below Spark-TTS and LLaSA. Against CosyVoice 3, under the same 0.5B LM scale, 25\,Hz frame rate, and DiT backbone, BiSAT-TTS achieves much lower error rates on every set with roughly one-eighth of its 1\,M-hour training set \cite{du2025cosyvoice3}, widening the LibriSpeech-PC WER margin from 0.11 points in reconstruction to 0.23 points in TTS. Comparing Tables\,\ref{tab:tokenizer_eval}, \ref{tab:tokenizer_understanding}, and \ref{tab:tts_results} clarifies how code structure affects AR modeling, notably on X-Codec2. Although X-Codec2 nears BiSAT in LibriSpeech-PC reconstruction WER (2.27\% vs.\ 2.15\%) at over twice the bitrate, its weak ASR, PR, and SID probes (Table\,\ref{tab:tokenizer_understanding}) show that AT-based reconstruction can achieve low error without cleanly structured codes. Consequently, both LLaSA-1B and LLaSA-3B still have much higher error and Fail rates than BiSAT-TTS on every set despite up to 6$\times$ the LM scale and twice the training data. Strong reconstruction can thus mask weakly structured codes that hinder AR generation, whereas BiSAT's lower error and Fail rates alongside Table\,\ref{tab:tokenizer_understanding}'s results indicate that preserving accessible semantic and acoustic structure yields code representations better suited for AR speech modeling.\looseness=-1

\textbf{Speaker Similarity.} BiSAT-TTS closely approaches CosyVoice 3's SIM-o on LibriSpeech-PC and trails only CosyVoice 3 on Seed-TTS test-en. The larger gap to CosyVoice 2 and 3 on test-zh may reflect their larger Mandarin training sets rather than a tokenizer bottleneck, consistent with BiSAT's leading SID and ASV scores in Table\,\ref{tab:tokenizer_understanding}.\looseness=-1

\textbf{Perceived Naturalness and Speech Quality.} Relative to CosyVoice 3, BiSAT-TTS achieves comparable UTMOSv2 scores on all three sets, whereas CosyVoice 2 and Spark-TTS score highest. Unlike LLaSA and Spark-TTS, which use AT waveform decoders \cite{ye2025llasa,wang2025sparktts}, BiSAT-TTS and CosyVoice condition flow-matching decoders on reference mel prompts. Because prompt-conditioned flow matching tends to preserve prompt acoustics and background noise \cite{wang24v_interspeech}, reference-free UTMOSv2 scores can be lower on in-the-wild Seed-TTS prompts than for AT decoders. On NISQA, which evaluates overall quality across noisiness, coloration, discontinuity, and loudness \cite{mittag2021nisqa}, both BiSAT-TTS variants still outperform CosyVoice 3, Spark-TTS, and both LLaSA models on all three sets. CosyVoice 2's strong scores on both metrics may also reflect its convolutional Transformer U-Net decoder rather than the DiT in CosyVoice 3 and BiSAT \cite{du2024cosyvoice2,du2025cosyvoice3}. Overall, BiSAT balances intelligibility, speaker similarity, naturalness, and speech quality well in practice.\looseness=-1

\subsubsection{Codebook Size Ablation}
\label{sssec:codebook_ablation}
Larger codebooks increase capacity but risk VQ collapse and expand the AR output vocabulary, making next-token prediction harder and typically demanding larger models and more training data. Comparing BiSAT-32k ($K=32768$, 375\,bps) and BiSAT-64k ($K=65536$, 400\,bps) across Tables\,\ref{tab:tokenizer_eval} and \ref{tab:tts_results}, maintaining $99.8\%$ training utilization at $64$\,k for $6.7\%$ higher bitrate shows that shared-index assignment across separate continuous spaces, together with spherical k-means initialization, EMA updates, and dead-code replacement, prevents collapse even in a large paired vocabulary. In Table\,\ref{tab:tokenizer_eval}, BiSAT-64k slightly improves LibriSpeech-PC UTMOSv2 scores at comparable WER and SIM while modestly lowering WER and CER on most Common Voice 27 sets except Japanese, indicating that $32$\,k suffices for clean speech whereas $64$\,k improves robustness on noisier in-the-wild speech. PolyShift edges out cosine on most BiSAT-32k reconstruction sets, while both schedules perform comparably at $64$\,k and in TTS. In Table\,\ref{tab:tts_results}, BiSAT-TTS-64k lowers Seed-TTS WER/CER, raises Seed-TTS UTMOSv2, and improves NISQA across all three sets, with comparable LibriSpeech-PC WER and nearly unchanged SIM-o. Since LM output-head size and sampling settings also differ, this reflects an end-to-end trade-off while showing that BiSAT scales effectively to a $64$\,k codebook without degrading AR modeling under the same LM scale and data.\looseness=-1

\section{Conclusion}
\label{sec:conclusion}
We proposed BiSAT, a unified speech tokenizer that jointly optimizes separate acoustic and semantic branches into a single code stream via shared-index assignment without fixed semantic teachers. Built on BiSAT, BiSAT-TTS pairs an AR text-speech LM with a conditional flow-matching acoustic decoder for zero-shot TTS. Experiments across reconstruction, understanding, and diagnostic analyses show that BiSAT preserves linguistic and acoustic structure, generalizes to in-the-wild and out-of-domain speech, and learns complementary paired codebooks. In zero-shot TTS, BiSAT-TTS matches or surpasses state-of-the-art AR baselines with less training data. Finally, these results show that reconstruction metrics alone do not capture accessible structure for AR language modeling, making frozen-code understanding probes essential complementary indicators of token quality.\looseness=-1

\section{Limitations and Future Work}
\label{sec:limitation}
Due to our limited training budget, we could not validate BiSAT at a larger scale, especially for the tokenizer, where both encoders are initialized from the English-pretrained NeMo SSL Conformer rather than a larger multilingual encoder. Combined with much less Mandarin and Japanese training data \cite{du2025cosyvoice3}, this may limit performance on those languages compared with CosyVoice 3, and scaling the text-speech LM also remains to be studied. Unlike end-to-end waveform codecs such as X-Codec2 that reconstruct waveforms directly from codes alone, BiSAT relies on a separate speaker-conditioned flow-matching decoder and vocoder.
Despite our diagnostic analysis and codebook-size ablation, this study evaluates BiSAT as a whole without controlled full-scale ablations of its individual architectural components. In future work, we plan to isolate these component contributions, explore direct end-to-end waveform reconstruction, and extend BiSAT to code-based RL post-training and speech-to-speech LMs.\looseness=-1

\bibliographystyle{IEEEtran}
\bibliography{refs}

@string{interspeech  = "Proc. Interspeech"}

@string{icassp       = "Proc. ICASSP"}

@string{slt          = "Proc. SLT"}

@string{asru         = "Proc. ASRU"}

@string{neurips      = "Proc. NeurIPS"}

@string{neuripsdgm   = "Proc. NeurIPS Workshop Deep Generative Models"}

@string{iclr         = "Proc. ICLR"}

@string{icml         = "Proc. ICML"}

@string{cvpr         = "Proc. CVPR"}

@string{iccv         = "Proc. ICCV"}

@string{acl          = "Proc. ACL"}

@string{aclfindings  = "Findings of ACL"}

@string{aaai         = "Proc. AAAI"}

@string{allerton     = "Proc. Allerton Conf. Commun., Control, Comput."}

@string{anlp         = "Proc. Annu. Meeting Assoc. Natural Lang. Process."}

@string{taslp        = "IEEE/ACM Trans. Audio, Speech, Lang. Process."}

@string{taslpold     = "IEEE Trans. Audio, Speech, Lang. Process."}

@string{tnnls        = "IEEE Trans. Neural Netw. Learn. Syst."}

@string{taffc        = "IEEE Trans. Affect. Comput."}

@string{jstsp        = "IEEE J. Sel. Topics Signal Process."}

@string{tmlr         = "Trans. Mach. Learn. Res."}

@string{lre          = "Lang. Resour. Eval."}

@article{wang2023neural,
  title={Neural Codec Language Models are Zero-Shot Text to Speech Synthesizers},
  author={Chen, Sanyuan and Wang, Chengyi and Wu, Yu and Zhang, Ziqiang and Zhou, Long and Liu, Shujie and Chen, Zhuo and Liu, Yanqing and Wang, Huaming and Li, Jinyu and He, Lei and Zhao, Sheng and Wei, Furu},
  journal=taslpold,
  volume={33},
  pages={705--718},
  year={2025},
  doi={10.1109/TASLPRO.2025.3530270}
}

@inproceedings{kim2024clamtts,
  title={{CL}a{M}-{TTS}: Improving Neural Codec Language Model for Zero-Shot Text-to-Speech},
  author={Kim, Jaehyeon and Lee, Keon and Chung, Seungjun and Cho, Jaewoong},
  booktitle=iclr,
  pages={50278--50301},
  year={2024},
  eprint={2404.02781},
  archivePrefix={arXiv},
  url={https://openreview.net/forum?id=ofzeypWosV}
}

@article{hu2026qwen3tts,
  title={Qwen3-TTS Technical Report},
  author={Hu, Hangrui and Zhu, Xinfa and He, Ting and Guo, Dake and Zhang, Bin and Wang, Xiong and Guo, Zhifang and Jiang, Ziyue and Hao, Hongkun and Guo, Zishan and Zhang, Xinyu and Zhang, Pei and Yang, Baosong and Xu, Jin and Zhou, Jingren and Lin, Junyang},
  journal={arXiv preprint arXiv:2601.15621},
  year={2026}
}

@article{du2024cosyvoice2,
  title={CosyVoice 2: Scalable Streaming Speech Synthesis with Large Language Models},
  author={Du, Zhihao and Wang, Yuxuan and Chen, Qian and Shi, Xian and Lv, Xiang and Zhao, Tianyu and Gao, Zhifu and Yang, Yexin and Gao, Changfeng and Wang, Hui and Yu, Fan and Liu, Huadai and Sheng, Zhengyan and Gu, Yue and Deng, Chong and Wang, Wen and Zhang, Shiliang and Yan, Zhijie and Zhou, Jingren},
  journal={arXiv preprint arXiv:2412.10117},
  year={2024}
}

@inproceedings{qu2024tokenflow,
  title={{TokenFlow}: Unified Image Tokenizer for Multimodal Understanding and Generation},
  author={Qu, Liao and Zhang, Huichao and Liu, Yiheng and Wang, Xu and Jiang, Yi and Gao, Yiming and Ye, Hu and Du, Daniel K. and Yuan, Zehuan and Wu, Xinglong},
  booktitle=cvpr,
  pages={2545--2555},
  year={2025},
  doi={10.1109/CVPR52734.2025.00243}
}

@inproceedings{vandenoord2017vqvae,
  title={Neural Discrete Representation Learning},
  author={van den Oord, Aaron and Vinyals, Oriol and Kavukcuoglu, Koray},
  booktitle=neurips,
  volume={30},
  eprint={1711.00937},
  archivePrefix={arXiv},
  year={2017}
}

@inproceedings{lipman2023flow,
  title={Flow Matching for Generative Modeling},
  author={Lipman, Yaron and Chen, Ricky T. Q. and Ben-Hamu, Heli and Nickel, Maximilian and Le, Matt},
  booktitle=iclr,
  year={2023},
  eprint={2210.02747},
  archivePrefix={arXiv},
  url={https://openreview.net/forum?id=PqvMRDCJT9t}
}

@inproceedings{liu2023flow,
  title={Flow Straight and Fast: Learning to Generate and Transfer Data with Rectified Flow},
  author={Liu, Xingchao and Gong, Chengyue and Liu, Qiang},
  booktitle=iclr,
  year={2023},
  eprint={2209.03003},
  archivePrefix={arXiv},
  url={https://openreview.net/forum?id=XVjTT1nw5z}
}

@inproceedings{baevski2020wav2vec2,
  title={wav2vec 2.0: A Framework for Self-Supervised Learning of Speech Representations},
  author={Baevski, Alexei and Zhou, Yuhao and Mohamed, Abdelrahman and Auli, Michael},
  booktitle=neurips,
  volume={33},
  pages={12449--12460},
  eprint={2006.11477},
  archivePrefix={arXiv},
  year={2020}
}

@article{hsu2021hubert,
  title={{HuBERT}: Self-Supervised Speech Representation Learning by Masked Prediction of Hidden Units},
  author={Hsu, Wei-Ning and Bolte, Benjamin and Tsai, Yao-Hung Hubert and Lakhotia, Kushal and Salakhutdinov, Ruslan and Mohamed, Abdelrahman},
  journal=taslp,
  volume={29},
  pages={3451--3460},
  year={2021},
}

@article{borsos2023audiolm,
  title={{AudioLM}: A Language Modeling Approach to Audio Generation},
  author={Borsos, Zal{\'a}n and Marinier, Rapha{\"e}l and Vincent, Damien and Kharitonov, Eugene and Pietquin, Olivier and Sharifi, Matt and Roblek, Dominik and Teboul, Olivier and Grangier, David and Tagliasacchi, Marco and Zeghidour, Neil},
  journal=taslp,
  volume={31},
  pages={2523--2533},
  year={2023},
  doi={10.1109/TASLP.2023.3288409}
}

@inproceedings{zhang2024speechtokenizer,
  title={{SpeechTokenizer}: Unified Speech Tokenizer for Speech Language Models},
  author={Zhang, Xin and Zhang, Dong and Li, Shimin and Zhou, Yaqian and Qiu, Xipeng},
  booktitle=iclr,
  pages={31798--31818},
  year={2024},
  eprint={2308.16692},
  archivePrefix={arXiv},
  url={https://openreview.net/forum?id=AF9Q8Vip84}
}

@article{defossez2023encodec,
  title={High Fidelity Neural Audio Compression},
  author={D{\'e}fossez, Alexandre and Copet, Jade and Synnaeve, Gabriel and Adi, Yossi},
  journal=tmlr,
  year={2023},
  eprint={2210.13438},
  archivePrefix={arXiv},
  url={https://openreview.net/forum?id=ivCd8z8zR2}
}

@article{chen2022wavlm,
  title={{WavLM}: Large-Scale Self-Supervised Pre-Training for Full Stack Speech Processing},
  author={Chen, Sanyuan and Wang, Chengyi and Chen, Zhengyang and Wu, Yu and Liu, Shujie and Chen, Zhuo and Li, Jinyu and Kanda, Naoyuki and Yoshioka, Takuya and Xiao, Xiong and Wu, Jian and Zhou, Long and Ren, Shuo and Qian, Yanmin and Qian, Yao and Wu, Jian and Zeng, Michael and Yu, Xiangzhan and Wei, Furu},
  journal=jstsp,
  volume={16},
  number={6},
  pages={1505--1518},
  year={2022},
  doi={10.1109/JSTSP.2022.3188113}
}

@inproceedings{siuzdak2024vocos,
  title={Vocos: Closing the Gap between Time-Domain and Fourier-Based Neural Vocoders for High-Quality Audio Synthesis},
  author={Siuzdak, Hubert},
  booktitle=iclr,
  pages={25719--25733},
  year={2024},
  eprint={2306.00814},
  archivePrefix={arXiv},
  url={https://openreview.net/forum?id=vY9nzQmQBw}
}

@article{su2024roformer,
  title={{RoFormer}: Enhanced Transformer with Rotary Position Embedding},
  author={Su, Jianlin and Ahmed, Murtadha and Lu, Yu and Pan, Shengfeng and Bo, Wen and Liu, Yunfeng},
  journal={Neurocomputing},
  volume={568},
  pages={127063},
  year={2024},
  doi={10.1016/j.neucom.2023.127063}
}

@article{chen2024valle2,
  title={{VALL-E 2}: Neural Codec Language Models are Human Parity Zero-Shot Text to Speech Synthesizers},
  author={Chen, Sanyuan and Liu, Shujie and Zhou, Long and Liu, Yanqing and Tan, Xu and Li, Jinyu and Zhao, Sheng and Qian, Yao and Wei, Furu},
  journal={arXiv preprint arXiv:2406.05370},
  eprint={2406.05370},
  archivePrefix={arXiv},
  year={2024}
}

@inproceedings{mehta2024matcha,
  title={{Matcha-TTS}: A Fast {TTS} Architecture with Conditional Flow Matching},
  author={Mehta, Shivam and Tu, Ruibo and Beskow, Jonas and Sz{\'e}kely, {\'E}va and Henter, Gustav Eje},
  booktitle=icassp,
  pages={11341--11345},
  year={2024},
}

@inproceedings{peebles2023dit,
    title={Scalable Diffusion Models with Transformers},
    author={Peebles, William and Xie, Saining},
    booktitle=iccv,
    year={2023}
}

@inproceedings{tang2022unified,
  title={Unified Speech-Text Pre-training for Speech Translation and Recognition},
  author={Tang, Yun and Gong, Hongyu and Dong, Ning and Wang, Changhan and Hsu, Wei-Ning and Gu, Jiatao and Baevski, Alexei and Li, Xian and Mohamed, Abdelrahman and Auli, Michael and Pino, Juan},
  booktitle=acl,
  pages={1488--1499},
  year={2022},
  doi={10.18653/v1/2022.acl-long.105}
}

@inproceedings{ao2022speecht5,
  title={{SpeechT5}: Unified-Modal Encoder-Decoder Pre-Training for Spoken Language Processing},
  author={Ao, Junyi and Wang, Rui and Zhou, Long and Wang, Chengyi and Ren, Shuo and Wu, Yu and Liu, Shujie and Ko, Tom and Li, Qing and Zhang, Yu and Wei, Zhihua and Qian, Yao and Li, Jinyu and Wei, Furu},
  booktitle=acl,
  pages={5723--5738},
  year={2022},
  doi={10.18653/v1/2022.acl-long.393}
}

@article{zeghidour2022soundstream,
  title={{SoundStream}: An End-to-End Neural Audio Codec},
  author={Zeghidour, Neil and Luebs, Alejandro and Omran, Ahmed and Skoglund, Jan and Tagliasacchi, Marco},
  journal=taslp,
  volume={30},
  pages={495--507},
  year={2022},
  doi={10.1109/TASLP.2021.3129994}
}

@inproceedings{razavi2019generating,
  title={Generating Diverse High-Fidelity Images with {VQ-VAE-2}},
  author={Razavi, Ali and van den Oord, Aaron and Vinyals, Oriol},
  booktitle=neurips,
  volume={32},
  eprint={1906.00446},
  archivePrefix={arXiv},
  year={2019}
}

@article{du2025cosyvoice3,
  title={{CosyVoice} 3: Towards In-the-wild Speech Generation via Scaling-up and Post-training},
  author={Du, Zhihao and Gao, Changfeng and Wang, Yuxuan and Yu, Fan and Zhao, Tianyu and Wang, Hao and Lv, Xiang and Wang, Hui and Ni, Chongjia and Shi, Xian and An, Keyu and Yang, Guanrou and Li, Yabin and Chen, Yanni and Gao, Zhifu and Chen, Qian and Gu, Yue and Chen, Mengzhe and Chen, Yafeng and Zhang, Shiliang and Wang, Wen and Ye, Jieping},
  journal={arXiv preprint arXiv:2505.17589},
  eprint={2505.17589},
  archivePrefix={arXiv},
  year={2025}
}

@article{du2024cosyvoice,
  title={{CosyVoice}: A Scalable Multilingual Zero-Shot Text-to-Speech Synthesizer based on Supervised Semantic Tokens},
  author={Du, Zhihao and Chen, Qian and Zhang, Shiliang and Hu, Kai and Lu, Heng and Yang, Yexin and Hu, Hangrui and Zheng, Siqi and Gu, Yue and Ma, Ziyang and Gao, Zhifu and Yan, Zhijie},
  journal={arXiv preprint arXiv:2407.05407},
  eprint={2407.05407},
  archivePrefix={arXiv},
  year={2024}
}

@article{jo2026lmspt,
  title={{LM-SPT}: {LM}-Aligned Semantic Distillation for Speech Tokenization},
  author={Jo, Daejin and Yun, Jeeyoung and Roh, Byungseok and Kim, Sungwoong},
  journal={IEEE Trans. Audio, Speech, Lang. Process.},
  volume={34},
  pages={3714--3727},
  year={2026},
  doi={10.1109/TASLPRO.2026.3705689}
}

@inproceedings{tishby1999information,
  title={The Information Bottleneck Method},
  author={Tishby, Naftali and Pereira, Fernando C. and Bialek, William},
  booktitle=allerton,
  pages={368--377},
  eprint={physics/0004057},
  archivePrefix={arXiv},
  year={1999}
}

@inproceedings{alemi2017deep,
  title={Deep Variational Information Bottleneck},
  author={Alemi, Alexander A. and Fischer, Ian and Dillon, Joshua V. and Murphy, Kevin},
  booktitle=iclr,
  year={2017},
  eprint={1612.00410},
  archivePrefix={arXiv},
  url={https://openreview.net/forum?id=HyxQzBceg}
}

@article{yang2025simplespeech2,
  title={{SimpleSpeech 2}: Towards Simple and Efficient Text-to-Speech with Flow-Based Scalar Latent Transformer Diffusion Models},
  author={Yang, Dongchao and Huang, Rongjie and Wang, Yuanyuan and Guo, Haohan and Chong, Dading and Liu, Songxiang and Wu, Xixin and Meng, Helen},
  journal=taslpold,
  year={2025},
}

@article{lee2025hierspeechpp,
  title={{HierSpeech++}: Bridging the Gap between Semantic and Acoustic Representation of Speech by Hierarchical Variational Inference for Zero-Shot Speech Synthesis},
  author={Lee, Sang-Hoon and Choi, Ha-Yeong and Kim, Seung-Bin and Lee, Seong-Whan},
  journal=tnnls,
  year={2025},
}

@article{zhou2025voxcpm,
  title={{VoxCPM}: Tokenizer-Free {TTS} for Context-Aware Speech Generation and True-to-Life Voice Cloning},
  author={Zhou, Yixuan and Zeng, Guoyang and Liu, Xin and Li, Xiang and Yu, Renjie and Wang, Ziyang and Ye, Runchuan and Sun, Weiyue and Gui, Jiancheng and Li, Kehan and Wu, Zhiyong and Liu, Zhiyuan},
  journal={arXiv preprint arXiv:2509.24650},
  eprint={2509.24650},
  archivePrefix={arXiv},
  year={2025}
}

@article{wang2025sparktts,
  title={{Spark-TTS}: An Efficient {LLM}-Based Text-to-Speech Model with Single-Stream Decoupled Speech Tokens},
  author={Wang, Xinsheng and Jiang, Mingqi and Ma, Ziyang and Zhang, Ziyu and Liu, Songxiang and Li, Linqin and Liang, Zheng and Zheng, Qixi and Wang, Rui and Feng, Xiaoqin and Bian, Weizhen and Ye, Zhen and Cheng, Sitong and Yuan, Ruibin and Zhao, Zhixian and Zhu, Xinfa and Pan, Jiahao and Xue, Liumeng and Zhu, Pengcheng and Chen, Yunlin and Li, Zhifei and Chen, Xie and Xie, Lei and Guo, Yike and Xue, Wei},
  journal={arXiv preprint arXiv:2503.01710},
  eprint={2503.01710},
  archivePrefix={arXiv},
  year={2025}
}

@inproceedings{shen2024naturalspeech2,
  title={{NaturalSpeech 2}: Latent Diffusion Models are Natural and Zero-Shot Speech and Singing Synthesizers},
  author={Shen, Kai and Ju, Zeqian and Tan, Xu and Liu, Yanqing and Leng, Yichong and He, Lei and Qin, Tao and Zhao, Sheng and Bian, Jiang},
  booktitle=iclr,
  pages={698--722},
  year={2024},
  eprint={2304.09116},
  archivePrefix={arXiv},
  url={https://openreview.net/forum?id=Rc7dAwVL3v}
}

@inproceedings{ju2024naturalspeech3,
  title={{NaturalSpeech 3}: Zero-Shot Speech Synthesis with Factorized Codec and Diffusion Models},
  author={Ju, Zeqian and Wang, Yuancheng and Shen, Kai and Tan, Xu and Xin, Detai and Yang, Dongchao and Liu, Yanqing and Leng, Yichong and Song, Kaitao and Tang, Siliang and Wu, Zhizheng and Qin, Tao and Li, Xiang-Yang and Ye, Wei and Zhang, Shikun and Bian, Jiang and He, Lei and Li, Jinyu and Zhao, Sheng},
  booktitle=icml,
  volume={235},
  pages={22605--22623},
  year={2024},
  eprint={2403.03100},
  archivePrefix={arXiv},
  url={https://proceedings.mlr.press/v235/ju24b.html}
}

@inproceedings{song2024ellav,
  title={{ELLA-V}: Stable Neural Codec Language Modeling with Alignment-Guided Sequence Reordering},
  author={Song, Yakun and Chen, Zhuo and Wang, Xiaofei and Ma, Ziyang and Chen, Xie},
  booktitle=aaai,
  volume={39},
  number={24},
  pages={25174--25182},
  year={2025},
  doi={10.1609/aaai.v39i24.34703}
}

@article{defossez2024moshi,
    title={Moshi: a speech-text foundation model for real-time dialogue},
    author={D{\'e}fossez, Alexandre and Mazar{\'e}, Laurent and Orsini, Manu and Royer, Am{\'e}lie and P{\'e}rez, Patrick and J{\'e}gou, Herv{\'e} and Grave, Edouard and Zeghidour, Neil},
    journal={arXiv preprint arXiv:2410.00037},
    eprint={2410.00037},
    archivePrefix={arXiv},
    year={2024}
}

@inproceedings{gong2025xytokenizer,
    title={{XY}-Tokenizer: Mitigating the Semantic-Acoustic Conflict in Low-Bitrate Speech Codecs},
    author={Gong, Yitian and Jin, Luozhijie and Chen, Kuangwei and Zhang, Dong and Deng, Ruifan and Yang, Xiaogui and Zhang, Xin and Fei, Zhaoye and Cheng, Qinyuan and Li, Shimin and Qiu, Xipeng},
    booktitle=acl,
    pages={9350--9369},
    year={2026},
    doi={10.18653/v1/2026.acl-long.423}
}

@inproceedings{hussein25_interspeech,
    title={{HASRD}: Hierarchical Acoustic and Semantic Representation Disentanglement},
    author={Hussein, Amir and Khurana, Sameer and Wichern, Gordon and Germain, Fran\c{c}ois G. and Le Roux, Jonathan},
    booktitle=interspeech,
    year={2025}
}

@inproceedings{eskimez2024e2tts,
    title={{E2 TTS}: Embarrassingly Easy Fully Non-Autoregressive Zero-Shot {TTS}},
    author={Eskimez, Sefik Emre and Wang, Xiaofei and Thakker, Manthan and Li, Canrun and Tsai, Chung-Hsien and Xiao, Zhen and Yang, Hemin and Zhu, Zirun and Tang, Min and Tan, Xu and Liu, Yanqing and Zhao, Sheng and Kanda, Naoyuki},
    booktitle=slt,
    pages={682--689},
    year={2024},
}

@inproceedings{chen2025f5tts,
    title={{F5-TTS}: A Fairytaler that Fakes Fluent and Faithful Speech with Flow Matching},
    author={Chen, Yushen and Niu, Zhikang and Ma, Ziyang and Deng, Keqi and Wang, Chunhui and Zhao, Jian and Yu, Kai and Chen, Xie},
    booktitle=acl,
    pages={6255--},
    year={2025}
}

@inproceedings{song2024tacolm,
    title={{TacoLM}: {GaTed} Attention Equipped Codec Language Model are Efficient Zero-Shot Text to Speech Synthesizers},
    author={Song, Yakun and Chen, Zhuo and Wang, Xiaofei and Ma, Ziyang and Yang, Guanrou and Chen, Xie},
    booktitle=interspeech,
    pages={4433--4437},
    year={2024},
    doi={10.21437/Interspeech.2024-1531}
}

@inproceedings{koizumi23_interspeech,
    title={{LibriTTS-R}: A Restored Multi-Speaker Text-to-Speech Corpus},
    author={Koizumi, Yuma and Zen, Heiga and Karita, Shigeki and Ding, Yifan and Yatabe, Kohei and Morioka, Nobuyuki and Bacchiani, Michiel and Zhang, Yu and Han, Wei and Bapna, Ankur},
    booktitle=interspeech,
    pages={5491--5495},
    year={2023}
}

@inproceedings{wang24b_interspeech,
    title={{GLOBE}: A High-quality English Corpus with Global Accents for Zero-shot Speaker Adaptive Text-to-Speech},
    author={Wang, Wenbin and Song, Yang and Jha, Sanjay},
    booktitle=interspeech,
    pages={1365--1369},
    year={2024},
    doi={10.21437/Interspeech.2024-70}
}

@inproceedings{he2024emilia,
    title={Emilia: An Extensive, Multilingual, and Diverse Speech Dataset for Large-Scale Speech Generation},
    author={He, Haorui and Shang, Zengqiang and Wang, Chaoren and Li, Xuyuan and Gu, Yicheng and Hua, Hua and Liu, Liwei and Yang, Chen and Li, Jiaqi and Shi, Peiyang and Wang, Yuancheng and Chen, Kai and Zhang, Pengyuan and Wu, Zhizheng},
    booktitle=slt,
    pages={885--890},
    year={2024}
}

@inproceedings{meng2025autoregressive,
    title={Autoregressive Speech Synthesis without Vector Quantization},
    author={Meng, Lingwei and Zhou, Long and Liu, Shujie and Chen, Sanyuan and Han, Bing and Hu, Shujie and Liu, Yanqing and Li, Jinyu and Zhao, Sheng and Wu, Xixin and Meng, Helen M. and Wei, Furu},
    booktitle=acl,
    year={2025},
    url={https://aclanthology.org/2025.acl-long.65/},
    doi={10.18653/v1/2025.acl-long.65},
    pages={1287--1300}
}

@inproceedings{woo2023convnextv2,
    title={{ConvNeXt} V2: Co-Designing and Scaling ConvNets with Masked Autoencoders},
    author={Woo, Sanghyun and Debnath, Shoubhik and Hu, Ronghang and Chen, Xinlei and Liu, Zhuang and Kweon, In So and Xie, Saining},
    booktitle=cvpr,
    pages={16133--16142},
    year={2023}
}

@inproceedings{peng2024voicecraft,
    title={{VoiceCraft}: Zero-Shot Speech Editing and Text-to-Speech in the Wild},
    author={Peng, Puyuan and Huang, Po-Yao and Li, Shang-Wen and Mohamed, Abdelrahman and Harwath, David},
    booktitle=acl,
    pages={12442--12462},
    year={2024},
    url={https://aclanthology.org/2024.acl-long.673/}
}

@inproceedings{peng2025voicestar,
    title={{VoiceStar}: Robust Zero-Shot Autoregressive {TTS} with Duration Control and Extrapolation},
    author={Peng, Puyuan and Zheng, Zhisheng and Li, Shang-Wen and Mohamed, Abdelrahman and Harwath, David},
    booktitle=aclfindings,
    pages={11737--11754},
    year={2026},
    doi={10.18653/v1/2026.findings-acl.570}
}

@inproceedings{ho2021classifierfree,
    title={Classifier-Free Diffusion Guidance},
    author={Ho, Jonathan and Salimans, Tim},
    booktitle=neuripsdgm,
    year={2021},
    eprint={2207.12598},
    archivePrefix={arXiv},
    url={https://openreview.net/forum?id=qw8AKxfYbI}
}

@inproceedings{dellalibera2025focalcodec,
    title={{FocalCodec}: Low-Bitrate Speech Coding via Focal Modulation Networks},
    author={Della Libera, Luca and Paissan, Francesco and Subakan, Cem and Ravanelli, Mirco},
    booktitle=neurips,
    pages={26926--26951},
    year={2025},
    doi={10.52202/085713-0801}
}

@inproceedings{ji2025wavtokenizer,
    title={{WavTokenizer}: an Efficient Acoustic Discrete Codec Tokenizer for Audio Language Modeling},
    author={Ji, Shengpeng and Jiang, Ziyue and Wang, Wen and Chen, Yifu and Fang, Minghui and Zuo, Jialong and Yang, Qian and Cheng, Xize and Wang, Zehan and Li, Ruiqi and Zhang, Ziang and Yang, Xiaoda and Huang, Rongjie and Jiang, Yidi and Chen, Qian and Zheng, Siqi and Zhao, Zhou},
    booktitle=iclr,
    pages={93809--93826},
    year={2025},
    eprint={2408.16532},
    archivePrefix={arXiv},
    url={https://openreview.net/forum?id=V8xX2n0F97}
}

@inproceedings{ye2025xcodec,
    title={Codec Does Matter: Exploring the Semantic Shortcoming of Codec for Audio Language Model},
    author={Ye, Zhen and Sun, Peiwen and Lei, Jiahe and Lin, Hongzhan and Tan, Xu and Dai, Zheqi and Kong, Qiuqiang and Chen, Jianyi and Pan, Jiahao and Liu, Qifeng and Guo, Yike and Xue, Wei},
    booktitle=aaai,
    year={2025}
}

@article{ye2025llasa,
    title={Llasa: Scaling Train-Time and Inference-Time Compute for Llama-based Speech Synthesis},
    author={Ye, Zhen and Zhu, Xinfa and Chan, Chi-Min and Wang, Xinsheng and Tan, Xu and Lei, Jiahe and Peng, Yi and Liu, Haohe and Jin, Yizhu and Dai, Zheqi and Lin, Hongzhan and Chen, Jianyi and Du, Xingjian and Xue, Liumeng and Chen, Yunlin and Li, Zhifei and Xie, Lei and Kong, Qiuqiang and Guo, Yike and Xue, Wei},
    journal={arXiv preprint arXiv:2502.04128},
    eprint={2502.04128},
    archivePrefix={arXiv},
    year={2025}
}

@inproceedings{graves2006ctc,
    title={Connectionist Temporal Classification: Labelling Unsegmented Sequence Data with Recurrent Neural Networks},
    author={Graves, Alex and Fern{\'a}ndez, Santiago and Gomez, Faustino and Schmidhuber, J{\"u}rgen},
    booktitle=icml,
    year={2006}
}

@inproceedings{gulati2020conformer,
    title={Conformer: Convolution-augmented Transformer for Speech Recognition},
    author={Gulati, Anmol and Qin, James and Chiu, Chung-Cheng and Parmar, Niki and Zhang, Yu and Yu, Jiahui and Han, Wei and Wang, Shibo and Zhang, Zhengdong and Wu, Yonghui and Pang, Ruoming},
    booktitle=interspeech,
    year={2020}
}

@inproceedings{fifty2025rotation,
    title={Restructuring Vector Quantization with the Rotation Trick},
    author={Fifty, Christopher and Junkins, Ronald G and Duan, Dennis and Iyengar, Aniketh and Liu, Jerry W and Amid, Ehsan and Thrun, Sebastian and R{\'e}, Christopher},
    booktitle=iclr,
    pages={19153--19188},
    year={2025},
    eprint={2410.06424},
    archivePrefix={arXiv},
    url={https://openreview.net/forum?id=nS2DBNydCC}
}

@inproceedings{peng2024owsmctc,
  title={{OWSM-CTC}: An Open Encoder-Only Speech Foundation Model for Speech Recognition, Translation, and Language Identification},
  author={Peng, Yifan and Sudo, Yui and Shakeel, Muhammad and Watanabe, Shinji},
  booktitle=acl,
  year={2024},
  pages={10192--10209},
  doi={10.18653/v1/2024.acl-long.549},
  url={https://aclanthology.org/2024.acl-long.549/}
}

@inproceedings{zhao2024bsq,
  title={Image and Video Tokenization with Binary Spherical Quantization},
  author={Zhao, Yue and Xiong, Yuanjun and Kr{\"a}henb{\"u}hl, Philipp},
  booktitle=iclr,
  pages={90844--90868},
  year={2025},
  eprint={2406.07548},
  archivePrefix={arXiv},
  url={https://arxiv.org/abs/2406.07548}
}

@inproceedings{loshchilov2019decoupled,
  title={Decoupled Weight Decay Regularization},
  author={Loshchilov, Ilya and Hutter, Frank},
  booktitle=iclr,
  year={2019},
  eprint={1711.05101},
  archivePrefix={arXiv},
  url={https://openreview.net/forum?id=Bkg6RiCqY7}
}

@inproceedings{wu2024codecsuperb,
    title={{Codec-SUPERB}: An In-Depth Analysis of Sound Codec Models},
    author={Wu, Haibin and Chung, Ho-Lam and Lin, Yi-Cheng and Wu, Yuan-Kuei and Chen, Xuanjun and Pai, Yu-Chi and Wang, Hsiu-Hsuan and Chang, Kai-Wei and Liu, Alexander H. and Lee, Hung-yi},
    booktitle=aclfindings,
    year={2024}
}

@article{kuchaiev2019nemo,
    title={{NeMo}: a toolkit for building {AI} applications using Neural Modules},
    author={Kuchaiev, Oleksii and Li, Jason and Nguyen, Huyen and Hrinchuk, Oleksii and Leary, Ryan and Ginsburg, Boris and Kriman, Samuel and Beliaev, Stanislav and Lavrukhin, Vitaly and Cook, Jack and others},
    journal={arXiv preprint arXiv:1909.09577},
    year={2019},
  url={https://arxiv.org/abs/1909.09577},
  eprint={1909.09577},
  archivePrefix={arXiv}
}

@article{moon2026deltatts,
  title={{DELTA-TTS}: Adapting Autoregressive Model into Diffusion Language Model for Text-to-Speech},
  author={Moon, Junwon and Lee, Yejin and Kim, Seungbeom and Ahn, Hoseong and Park, Sewoong and Kim, Heeseung and Shim, Kyuhong},
  journal={arXiv preprint arXiv:2607.04140},
  year={2026},
  url={https://arxiv.org/abs/2607.04140},
  eprint={2607.04140},
  archivePrefix={arXiv}
}

@misc{minicpm2025minicpm4,
  title={{MiniCPM4}: Ultra-Efficient {LLMs} on End Devices},
  author={{MiniCPM Team}},
  year={2025},
  howpublished={\url{https://huggingface.co/openbmb/MiniCPM4-0.5B}},
  eprint={2506.07900},
  archivePrefix={arXiv},
  url={https://arxiv.org/abs/2506.07900},
  note={Technical report: arXiv:2506.07900}
}

@inproceedings{park2019specaugment,
  title={{SpecAugment}: A Simple Data Augmentation Method for Automatic Speech Recognition},
  author={Park, Daniel S. and Chan, William and Zhang, Yu and Chiu, Chung-Cheng and Zoph, Barret and Cubuk, Ekin D. and Le, Quoc V.},
  booktitle=interspeech,
  year={2019},
  pages={2613--2617},
  doi={10.21437/Interspeech.2019-2680}
}

@inproceedings{wang2023campplus,
  title={{CAM++}: A Fast and Efficient Network for Speaker Verification Using Context-Aware Masking},
  author={Wang, Hui and Zheng, Siqi and Chen, Yafeng and Cheng, Luyao and Chen, Qian},
  booktitle=interspeech,
  year={2023},
  pages={5301--5305},
  doi={10.21437/Interspeech.2023-1513}
}

@inproceedings{shi2021aishell3,
  title={{AISHELL-3}: A Multi-Speaker Mandarin {TTS} Corpus},
  author={Shi, Yao and Bu, Hui and Xu, Xin and Zhang, Shaoji and Li, Ming},
  booktitle=interspeech,
  year={2021},
  pages={2756--2760}
}

@misc{yamagishi2019vctk,
  title={{CSTR VCTK} Corpus: English Multi-speaker Corpus for {CSTR} Voice Cloning Toolkit (version 0.92)},
  author={Yamagishi, Junichi and Veaux, Christophe and MacDonald, Kirsten},
  howpublished={University of Edinburgh, Centre for Speech Technology Research},
  year={2019},
  doi={10.7488/ds/2645}
}

@inproceedings{li2024yodas,
  title={{YODAS}: {Youtube}-Oriented Dataset for Audio and Speech},
  author={Li, Xinjian and Takamichi, Shinnosuke and Saeki, Takaaki and Chen, William and Shiota, Sayaka and Watanabe, Shinji},
  booktitle=asru,
  pages={1--8},
  year={2023},
  doi={10.1109/ASRU57964.2023.10389689}
}

@inproceedings{oliveira2023cmltts,
  title={{CML-TTS}: A Multilingual Dataset for Speech Synthesis in Low-Resource Languages},
  author={Oliveira, Frederico S. and Casanova, Edresson and {C\^andido J\'unior}, Arnaldo and Soares, Anderson S. and {Galv\~ao Filho}, Arlindo R.},
  booktitle={Proc. TSD},
  pages={188--199},
  year={2023},
  doi={10.1007/978-3-031-40498-6_17}
}

@inproceedings{puchtler2021hui,
  title={{HUI-Audio-Corpus-German}: A High Quality {TTS} Dataset},
  author={Puchtler, Pascal and Wirth, Johannes and Peinl, Ren\'e},
  booktitle={Proc. KI},
  pages={204--216},
  year={2021},
  doi={10.1007/978-3-030-87626-5_15}
}

@article{takamichi2019jvs,
  title={{JVS} Corpus: Free Japanese Multi-speaker Voice Corpus},
  author={Takamichi, Shinnosuke and Mitsui, Kentaro and Saito, Yuki and Koriyama, Tomoki and Tanji, Naoko and Saruwatari, Hiroshi},
  journal={arXiv preprint arXiv:1908.06248},
  year={2019},
  url={https://arxiv.org/abs/1908.06248},
  eprint={1908.06248},
  archivePrefix={arXiv}
}

@misc{ye2025xcodec2release,
  title={{X-Codec-2.0}},
  author={Ye, Zhen and others},
  year={2025},
  howpublished={\url{https://github.com/zhenye234/X-Codec-2.0}},
  note={Model checkpoint: \url{https://huggingface.co/HKUSTAudio/xcodec2}}
}

@inproceedings{yang2021superb,
    title={{SUPERB}: Speech Processing Universal {PERformance} Benchmark},
    author={Yang, Shu-wen and Chi, Po-Han and Chuang, Yung-Sung and Lai, Cheng-I Jeff and Lakhotia, Kushal and Lin, Yist Y and Liu, Andy T and Shi, Jiatong and Chang, Xuankai and Lin, Guan-Ting and others},
    booktitle=interspeech,
    year={2021}
}

@inproceedings{shi2023mlsuperb,
    title={{ML-SUPERB}: Multilingual Speech Universal {PERformance} Benchmark},
    author={Shi, Jiatong and Berrebbi, Dan and Chen, William and Chung, Ho-Lam and Hu, En-Pei and Huang, Wei Ping and Chang, Xuankai and Li, Shang-Wen and Mohamed, Abdelrahman and Lee, Hung-yi and Watanabe, Shinji},
    booktitle=interspeech,
    year={2023}
}

@inproceedings{panayotov2015librispeech,
    title={{Librispeech}: An {ASR} corpus based on public domain audio books},
    author={Panayotov, Vassil and Chen, Guoguo and Povey, Daniel and Khudanpur, Sanjeev},
    booktitle=icassp,
    year={2015}
}

@inproceedings{nagrani2017voxceleb,
    title={{VoxCeleb}: A Large-Scale Speaker Identification Dataset},
    author={Nagrani, Arsha and Chung, Joon Son and Zisserman, Andrew},
    booktitle=interspeech,
    year={2017}
}

@article{busso2008iemocap,
    title={{IEMOCAP}: Interactive emotional dyadic motion capture database},
    author={Busso, Carlos and Bulut, Murtaza and Lee, Chi-Chun and Kazemzadeh, Abe and Mower, Emily and Kim, Samuel and Chang, Jeannette N and Lee, Sungbok and Narayanan, Shrikanth S},
    journal=lre,
    volume={42},
    pages={335--359},
    year={2008}
}

@article{cao2014cremad,
    title={{CREMA-D}: Crowd-sourced Emotional Multimodal Actors Dataset},
    author={Cao, Houwei and Cooper, David G and Keutmann, Michael K and Gur, Ruben C and Nenkova, Ani and Verma, Ragini},
    journal=taffc,
    volume={5},
    number={4},
    pages={377--390},
    year={2014}
}

@article{livingstone2018ravdess,
    title={The {Ryerson} Audio-Visual Database of Emotional Speech and Song ({RAVDESS})},
    author={Livingstone, Steven R and Russo, Frank A},
    journal={PLoS ONE},
    volume={13},
    number={5},
    pages={e0196391},
    year={2018}
}

@inproceedings{zhou2021esd,
    title={Seen and Unseen Emotional Style Transfer for Voice Conversion with a New Emotional Speech Dataset},
    author={Zhou, Kun and Sisman, Berrak and Liu, Rui and Li, Haizhou},
    booktitle=icassp,
    year={2021}
}

@inproceedings{yin2023reazonspeech,
    title={{ReazonSpeech}: A Free and Massive Corpus for {Japanese} {ASR}},
    author={Yin, Yue and Mori, Daijiro and Fujimoto, Seiji},
    booktitle=anlp,
    year={2023},
  url={https://anlp.jp/proceedings/annual_meeting/2023/pdf_dir/A5-3.pdf}
}

@inproceedings{conneau2023fleurs,
    title={{FLEURS}: Few-shot Learning Evaluation of Universal Representations of Speech},
    author={Conneau, Alexis and Ma, Min and Khanuja, Simran and Zhang, Yu and Axelrod, Vera and Dalmia, Siddharth and Riesa, Jason and Rivera, Clara and Bapna, Ankur},
    booktitle=slt,
    year={2023}
}

@article{chen2026wavtts,
    title={{WavTTS}: Towards High-Quality Zero-Shot {TTS} via Direct Raw Waveform Modeling},
    author={Chen, Wenxi and Jia, Dongya and Chen, Yushen and Niu, Zhikang and Liang, Yuzhe and Li, Xiquan and Yan, Ruiqi and Ma, Ziyang and Yang, Guanrou and Chen, Sanyuan and Wang, Yue and Chen, Zhuo and Yu, Kai and Chen, Xie},
    journal={arXiv preprint arXiv:2606.03455},
    year={2026},
  url={https://arxiv.org/abs/2606.03455},
  eprint={2606.03455},
  archivePrefix={arXiv}
}

@inproceedings{mentzer2024fsq,
    title={Finite Scalar Quantization: {VQ-VAE} Made Simple},
    author={Mentzer, Fabian and Minnen, David and Agustsson, Eirikur and Tschannen, Michael},
    booktitle=iclr,
    year={2024},
  url={https://arxiv.org/abs/2309.15505},
  eprint={2309.15505},
  archivePrefix={arXiv}
}

@inproceedings{niu2026semanticvae,
    title={{Semantic-VAE}: Semantic-Alignment Latent Representation for Better Speech Synthesis},
    author={Niu, Zhikang and others},
    booktitle=interspeech,
    year={2026}
}

@inproceedings{pratap2020mls,
    title={{MLS}: A Large-Scale Multilingual Dataset for Speech Research},
    author={Pratap, Vineel and Xu, Qiantong and Sriram, Anuroop and Synnaeve, Gabriel and Collobert, Ronan},
    booktitle=interspeech,
    year={2020}
}

@inproceedings{zhang2022wenetspeech,
    title={{WenetSpeech}: A 10000+ Hours Multi-domain Mandarin Corpus for Speech Recognition},
    author={Zhang, Binbin and Lv, Hang and Guo, Pengcheng and Shao, Qijie and Yang, Chao and Xie, Lei and Xu, Xin and Bu, Hui and Chen, Xiaoyu and Zeng, Chenchen and Wu, Di and Peng, Zhendong},
    booktitle=icassp,
    year={2022}
}

@inproceedings{meister2023librispeechpc,
    title={{LibriSpeech-PC}: Benchmark for Evaluation of Punctuation and Capitalization Capabilities of End-to-end {ASR} Models},
    author={Meister, Aleksandr and Novikov, Matvei and Karpov, Nikolay and Bakhturina, Evelina and Lavrukhin, Vitaly and Ginsburg, Boris},
    booktitle=asru,
    year={2023}
}

@inproceedings{baba2024utmosv2,
    title={The {T05} System for the {VoiceMOS} Challenge 2024: Transfer Learning from Deep Image Classifier to Naturalness {MOS} Prediction of High-Quality Synthetic Speech},
    author={Baba, Kaito and Nakata, Wataru and Saito, Yuki and Saruwatari, Hiroshi},
    booktitle=slt,
    year={2024}
}

@inproceedings{radford2023whisper,
    title={Robust Speech Recognition via Large-Scale Weak Supervision},
    author={Radford, Alec and Kim, Jong Wook and Xu, Tao and Brockman, Greg and McLeavey, Christine and Sutskever, Ilya},
    booktitle=icml,
    pages={28492--28518},
    year={2023},
  url={https://proceedings.mlr.press/v202/radford23a.html},
  eprint={2212.04356},
  archivePrefix={arXiv}
}

@inproceedings{valk2021voxlingua107,
    title={{VoxLingua107}: A Dataset for Spoken Language Recognition},
    author={Valk, J{\"o}rgen and Alum{\"a}e, Tanel},
    booktitle=slt,
    pages={652--658},
    year={2021}
}

@IEEEtranBSTCTL{CAST:BSTcontrol,
  CTLuse_article_number    = "yes",
  CTLuse_paper             = "yes",
  CTLuse_url               = "no",
  CTLuse_forced_etal       = "yes",
  CTLmax_names_forced_etal = "6",
  CTLnames_show_etal       = "1",
  CTLdash_repeated_names   = "no"
}

@article{anastassiou2024seedtts,
  title={Seed-{TTS}: A Family of High-Quality Versatile Speech Generation Models},
  author={Anastassiou, Philip and Chen, Jiawei and Chen, Jitong and Chen, Yuanzhe and Chen, Zhuo and Chen, Ziyi and Cong, Jian and Deng, Lelai and Ding, Chuang and Gao, Lu and others},
  journal={arXiv preprint arXiv:2406.02430},
  year={2024},
  url={https://arxiv.org/abs/2406.02430},
  eprint={2406.02430},
  archivePrefix={arXiv}
}

@inproceedings{gao2022paraformer,
  title={Paraformer: Fast and Accurate Parallel Transformer for Non-autoregressive End-to-End Speech Recognition},
  author={Gao, Zhifu and Zhang, Shiliang and McLoughlin, Ian and Yan, Zhijie},
  booktitle=interspeech,
  pages={2063--2067},
  year={2022},
  doi={10.21437/Interspeech.2022-9996}
}

@inproceedings{ardila2020commonvoice,
  title={Common Voice: A Massively-Multilingual Speech Corpus},
  author={Ardila, Rosana and Branson, Megan and Davis, Kelly and Kohler, Michael and Meyer, Josh and Henretty, Michael and Morais, Reuben and Saunders, Lindsay and Tyers, Francis and Weber, Gregor},
  booktitle={Proc. LREC},
  pages={4218--4222},
  year={2020},
  url={https://aclanthology.org/2020.lrec-1.520/},
  eprint={1912.06670},
  archivePrefix={arXiv}
}

@inproceedings{wang24v_interspeech,
  title={An Investigation of Noise Robustness for Flow-Matching-Based Zero-Shot {TTS}},
  author={Wang, Xiaofei and Eskimez, Sefik Emre and Thakker, Manthan and Yang, Hemin and Zhu, Zirun and Tang, Min and Xia, Yufei and Li, Jinzhu and Zhao, Sheng and Li, Jinyu and Kanda, Naoyuki},
  booktitle=interspeech,
  pages={687--691},
  year={2024},
  doi={10.21437/Interspeech.2024-1336}
}

@inproceedings{kajarekar2003speaker,
  title={Speaker Recognition Using Prosodic and Lexical Features},
  author={Kajarekar, S. and Ferrer, L. and Venkataraman, A. and Sonmez, K. and Shriberg, E. and Stolcke, A. and Bratt, H. and Gadde, R. R.},
  booktitle=asru,
  pages={19--24},
  year={2003},
  doi={10.1109/ASRU.2003.1318397}
}

@book{gersho1992vector,
  title={Vector Quantization and Signal Compression},
  author={Gersho, Allen and Gray, Robert M.},
  publisher={Kluwer Academic Publishers},
  address={Boston, MA, USA},
  year={1992},
  doi={10.1007/978-1-4615-3626-0}
}

@article{glielmo2022ranking,
  title={Ranking the Information Content of Distance Measures},
  author={Glielmo, Aldo and Zeni, Claudio and Cheng, Bingqing and Cs{\'a}nyi, G{\'a}bor and Laio, Alessandro},
  journal={PNAS Nexus},
  volume={1},
  number={2},
  pages={pgac039},
  year={2022},
  doi={10.1093/pnasnexus/pgac039}
}

@inproceedings{chen2016xgboost,
  title={{XGBoost}: A Scalable Tree Boosting System},
  author={Chen, Tianqi and Guestrin, Carlos},
  booktitle={Proc. ACM SIGKDD},
  pages={785--794},
  year={2016},
  doi={10.1145/2939672.2939785}
}

@article{lundberg2020local,
  title={From Local Explanations to Global Understanding with Explainable {AI} for Trees},
  author={Lundberg, Scott M. and Erion, Gabriel and Chen, Hugh and DeGrave, Alex and Prutkin, Jordan M. and Nair, Bala and Katz, Ronit and Himmelfarb, Jonathan and Bansal, Nisha and Lee, Su-In},
  journal={Nature Mach. Intell.},
  volume={2},
  number={1},
  pages={56--67},
  year={2020},
  doi={10.1038/s42256-019-0138-9}
}

@inproceedings{mittag2021nisqa,
  title={{NISQA}: A Deep {CNN}-Self-Attention Model for Multidimensional Speech Quality Prediction with Crowdsourced Datasets},
  author={Mittag, Gabriel and Naderi, Babak and Chehadi, Assmaa and M{\"o}ller, Sebastian},
  booktitle=interspeech,
  pages={2127--2131},
  year={2021},
  doi={10.21437/Interspeech.2021-299}
}

\end{document}